\documentclass[twocolumn]{article}

\usepackage[T1]{fontenc}
\usepackage[utf8]{inputenc}
\usepackage{lmodern}
\usepackage[a4paper, top=2.2cm, bottom=2.2cm, left=1.8cm, right=1.8cm]{geometry}
\usepackage{amsmath,amssymb,bm}
\usepackage{graphicx}
\usepackage{xcolor}
\usepackage{booktabs}
\usepackage{multirow}
\usepackage{tabularx}
\usepackage{algorithm}
\usepackage{algpseudocode}
\usepackage{tikz}
\usepackage{pgfplots}
\pgfplotsset{compat=1.18}
\usepackage{listings}
\usepackage[colorlinks=true, linkcolor=nodered!70!black,
            citecolor=nodeblue!80!black, urlcolor=nodeblue!80!black]{hyperref}
\usepackage{cleveref}
\usepackage{microtype}
\usepackage{caption}
\usepackage{subcaption}
\usepackage[numbers,sort&compress]{natbib}
\usepackage{authblk}
\usepackage{orcidlink}

\usetikzlibrary{arrows.meta, positioning, shapes.geometric, fit, backgrounds,
                decorations.pathreplacing, calc}

\definecolor{nodeblue}{RGB}{52,120,190}
\definecolor{nodegreen}{RGB}{56,142,60}
\definecolor{nodeorange}{RGB}{230,81,0}
\definecolor{nodegray}{RGB}{97,97,97}
\definecolor{nodered}{RGB}{183,28,28}
\definecolor{lightgray}{RGB}{240,240,240}
\definecolor{revisionred}{RGB}{204,0,0}

\newif\ifrevisionsvisible
\revisionsvisiblefalse
\newcommand{\revised}[1]{%
  \ifrevisionsvisible\textcolor{revisionred}{#1}\else#1\fi}

\title{\textbf{PDE-Agents: An LLM-Orchestrated Multi-Agent\\
       Framework for Automated Finite Element Simulations\\
       with Knowledge Graph-Augmented Reasoning}}

\author[1]{Sayan Adhikari\,\orcidlink{0000-0002-9316-3292}}
\author[1]{Gulshan Noorsumar\,\orcidlink{0000-0002-6718-4508}}
\author[1]{\O yvind Jensen\,\orcidlink{0000-0003-4353-9359}}
\affil[1]{MatPro, Institute for Energy Technology (IFE), Kjeller, Norway}

\date{Preprint -- \today}

\hypersetup{
  pdftitle={PDE-Agents: An LLM-Orchestrated Multi-Agent Framework for
            Automated Finite Element Simulations with Knowledge
            Graph-Augmented Reasoning},
  pdfauthor={Sayan Adhikari, Gulshan Noorsumar, Øyvind Jensen},
  pdfsubject={LLM agents, finite element method, knowledge graphs},
  pdfkeywords={LLM, FEM, knowledge graph, GraphRAG, multi-agent systems}
}

\begin{document}
\maketitle

\let\thefootnote\relax\footnotetext{%
  \textit{Corresponding author:} Sayan Adhikari
  (\texttt{sayan.adhikari@ife.no}).\\
  \textit{Code:} \url{https://github.com/MatPro-IFE/pde-agents}}

\begin{abstract}
We present \textit{PDE-Agents}, a multi-agent ecosystem that automates the
full lifecycle of partial differential equation (PDE) / finite element method
(FEM) simulations through natural-language interaction.  The system combines three specialist large language model (LLM)
agents—Simulation, Analytics, and Database—orchestrated via a LangGraph
supervisor graph, with a locally-deployed open-source LLM stack
(Qwen3-Coder-Next, Llama~4~Scout) running on dual NVIDIA RTX PRO 6000
Blackwell GPUs ($\approx$196\,GB combined VRAM).  The architecture is
model-agnostic; we report cross-model validation across two
generations of open-source LLMs.
A central novelty is the integration of a \emph{GraphRAG} knowledge base
(Neo4j + 768-dimensional vector embeddings) that enriches agents with curated
material properties, known failure patterns, and prior run lineage.
We report seven empirical contributions: (i) a formal verification and
validation (V\&V) study confirming second-order spatial convergence (\(\mathcal{O}(h^2)\))
for all three benchmark cases on the heat equation solver; (ii) a
three-way ablation study across 50 benchmark tasks with a frozen KG,
comparing KG~On, KG~Off, and KG~Smart,
showing that KG~Smart achieves 100\% success \emph{and} the highest
output quality (physics score 0.933 vs.\ 0.853 for KG~Off,
 MPF 0.926 vs.\ 0.796), and that our \emph{KG Smart}
integration, combining warm-start injection with lazy conditional
retrieval, matches KG~Off reliability while maximising output fidelity;
(iii) a novel-material experiment using three fictional materials
whose properties exist only in the KG, where KG~Smart achieves
near-perfect material property fidelity (MPF\,=\,1.00) versus 0.34 for the
KG-free baseline; (iv) a failure analysis tracing KG~On's 3 systematic failures to budget
exhaustion and timeout, establishing warm-start injection as the dominant
factor in KG~Smart's reliability advantage;
(v) an adaptive KG decision framework (Algorithm~1)
that selects the optimal retrieval mode per task;
(vi) production-scale agent quality
metrics from 1{,}369 real simulation runs showing 97.8\% overall success
and \revised{an 85.4\%} first-try rate; and (vii) a controlled 100-task KG growth experiment showing
that accumulated run history yields a difficulty-dependent quality
gain, with hard-task MPF improving by 8.8\% between passes while
easy and novel tasks remain at ceiling.  All code, models, and evaluation artifacts
are released openly.  Taken together, these results point to
\emph{integration pattern}, rather than knowledge content, as what
decides whether GraphRAG augmentation helps or hinders an LLM agent, and
they yield concrete design principles for autonomous FEM simulation
assistants.

\end{abstract}

\section{Introduction}
\label{sec:intro}

Finite element method simulations are central to engineering analysis across
structural mechanics, heat transfer, fluid dynamics, and electromagnetics.
Yet the gap between a domain expert's intent—``what if I use aluminium instead
of steel?''—and a running, verified simulation involves a sequence of error-prone
manual steps: geometry creation, boundary condition specification, solver
parameter selection, result interpretation, and iterative debugging.

The recent emergence of \emph{reasoning-capable} LLMs \citep{brown2020gpt3,
wei2022cot, yao2023react} has opened the possibility of agents that can
autonomously traverse this workflow.  Prior work on ``AI for scientific
computing'' has largely focused on surrogate modelling
\citep{lagaris1998neural, raissi2019pinn, karniadakis2021physics}, operator
learning \citep{li2021fno}, or dataset-specific fine-tuning
\citep{jumper2021alphafold}.  Much less attention has been paid to
\emph{procedural automation}—using LLMs to \emph{set up and manage} solvers
rather than to replace them.

Meanwhile, the \emph{retrieval-augmented generation} (RAG) paradigm
\citep{lewis2020rag} has demonstrated that grounding LLM outputs with
authoritative external knowledge dramatically reduces hallucination.  The
graph-structured variant, GraphRAG \citep{edge2024graphrag}, further enables
multi-hop reasoning over relational knowledge—particularly relevant in
engineering contexts where material properties, mesh requirements, and solver
stability rules form a rich, interconnected knowledge base.

\textbf{In this paper we make the following contributions:}
\begin{enumerate}
  \item We design and implement \textit{PDE-Agents}, a complete, containerised
        multi-agent ecosystem for automated FEM simulation
        (\cref{sec:system}).
  \item We describe a novel \emph{GraphRAG integration pattern} that encodes
        material properties, known failure modes, and prior run lineage as a
        Neo4j knowledge graph with Hierarchical Navigable Small World
        (HNSW) vector-indexed embeddings \citep{malkov2020hnsw}
        (\cref{sec:kg}).
  \item We conduct a rigorous V\&V study that confirms $\mathcal{O}(h^2)$
        spatial convergence for three closed-form benchmark cases, validating
        the numerical kernel against analytical solutions
        (\cref{sec:vv}).
  \item We run a three-way ablation study (KG~On, KG~Off, KG~Smart)
        over 50 tasks with a frozen knowledge graph.  KG-enabled modes
        produce higher-quality simulation outputs (physics score,
        material property fidelity) than the KG-free baseline, and
        \emph{KG~Smart} reaches 100\% success \emph{and} the highest
        output quality (\cref{sec:ablation}).
  \item We introduce a \emph{novel-material experiment} using three
        fictional materials absent from any LLM training corpus.
        KG Smart attains ${\approx}\,2.9\times$ the material property
        fidelity of the KG-free baseline, which fabricates physically
        incorrect properties (\cref{sec:novidium}).
  \item We trace KG~On's 3 systematic failures to budget exhaustion
        and timeout, revealing that warm-start injection, not lazy
        retrieval, is the dominant factor in KG~Smart's reliability
        advantage (\cref{sec:failure-analysis}).
  \item We report production-scale agent quality metrics derived from
        1{,}369 real simulation runs, together with a controlled
        100-task KG growth experiment in which the quality gain proves
        difficulty-dependent: hard-task MPF improves by 8.8\% as the KG
        accumulates run history (\cref{sec:metrics}).
\end{enumerate}

\section{Related Work}
\label{sec:related}

\paragraph{LLM agents for scientific computing.}
ChemCrow \citep{bran2023chemcrow} demonstrated that equipping an LLM with
chemistry tools (RDKit, reaction planners) enables autonomous laboratory-scale
reasoning.  SciAgent \citep{ma2024sciagent} and similar systems extend this
to broader scientific question answering.  Our work targets \emph{numerical
simulation} rather than knowledge retrieval, requiring tool-call pipelines that
produce and verify deterministic numerical outputs.

\paragraph{Autonomous FEM systems.}
FEniCS \citep{logg2012fenics} and its successor DOLFINx/FEniCSx
\citep{barrata2023dolfinx} provide Python-scriptable FEM solvers that are
amenable to LLM-driven automation.  Early work on ``simulation copilots''
\citep{bauer2023llm, jiang2024llm} shows that LLMs can generate solver
scripts from natural language with moderate success, but these approaches are
single-turn and lack closed-loop debugging.  More recently, ALL-FEM
\citep{deotale2026allfem} fine-tunes LLMs on 1{,}000+ verified FEniCS scripts
and embeds them in a multi-agent workflow achieving 71.8\% code-level success
on multiphysics benchmarks, and FEABench \citep{mudur2025feabench} provides a
systematic evaluation using COMSOL Multiphysics (88\% executable API calls).
PDE-Agents differs by using unmodified open-source LLMs with
knowledge-graph augmentation rather than domain-specific fine-tuning, and by
providing a controlled ablation of retrieval strategies with physics-aware
quality metrics.

\paragraph{Multi-agent orchestration.}
LangGraph \citep{langgraph2024} implements graph-structured agent workflows
that support conditional routing and persistent state, enabling the supervisor
pattern we adopt.  AutoGen \citep{wu2023autogen} and CrewAI \citep{crewai2024}
offer related frameworks; we choose LangGraph for its explicit
state graph semantics and tight LangChain integration.

\paragraph{Retrieval-augmented generation for engineering.}
RAG \citep{lewis2020rag} has been applied to engineering documentation
retrieval \citep{liao2023rag}, code generation \citep{gao2023rag}, and more
recently to simulation parameter suggestion \citep{yang2024rag}.  GraphRAG
\citep{edge2024graphrag} adds structured relational traversal.
Corrective RAG (CRAG) \citep{yan2024crag} introduces adaptive retrieval
decisions based on document relevance scoring, and AriGraph
\citep{arigraph2024} demonstrates episodic knowledge-graph memory for LLM
agents.  Our \emph{KG Smart} pattern draws on both: warm-start injection
from similar past runs (episodic memory) with lazy conditional tool
invocation (corrective retrieval).  To our knowledge, ours is the first
system to combine GraphRAG with a live, iterative FEM simulation loop.

\paragraph{Knowledge graphs for materials science.}
Materials-focused knowledge graphs (e.g.\ MatKG \citep{venugopal2022matkg},
OPTIMADE \citep{andersen2021optimade}) encode material properties in RDF/OWL
formats.  \citet{buehler2024mechgpt} demonstrates that retrieval-augmented
ontological KG strategies outperform flat RAG for materials design by
capturing mechanistic relationships between concepts.
Our Neo4j KG is lighter-weight and purpose-built for run-time
agent queries, prioritising latency over completeness.

\section{System Architecture}
\label{sec:system}

\begin{figure*}[htbp]
  \centering
  \resizebox{\textwidth}{!}{
%
%
%
%
%
\begin{tikzpicture}[
  font=\small,
  >=Latex,
  every node/.style={align=center},
  usrnd/.style={
    draw=orange!60, fill=yellow!18, rounded corners=16pt,
    thick, minimum width=3.6cm, minimum height=0.76cm,
    font=\small\bfseries},
  supnd/.style={
    draw=nodeblue, fill=nodeblue!15, rounded corners=5pt,
    very thick, minimum width=13.4cm, minimum height=0.9cm,
    font=\small\bfseries},
  agentnd/.style 2 args={
    draw=#1!88, fill=#1!10, rounded corners=5pt, thick,
    minimum width=3.4cm, minimum height=1.26cm, font=\small},
  svcnd/.style 2 args={
    draw=#1!78, fill=#1!8, rounded corners=4pt, thick,
    minimum width=3.4cm, minimum height=1.15cm, font=\small},
  storend/.style={
    draw=nodegray!60, fill=nodegray!7, rounded corners=4pt, thick,
    minimum width=3.4cm, minimum height=1.05cm, font=\small},
  llmnd/.style={
    draw=nodeblue!58, fill=nodeblue!7, rounded corners=5pt, very thick,
    minimum width=14.0cm, minimum height=1.05cm, font=\footnotesize},
  lbox/.style={
    draw=#1!42, fill=#1!4, rounded corners=8pt, thick, dashed,
    inner sep=0.42cm},
  arr/.style   ={->,    thick,      color=#1},
  biarr/.style ={<->,   thick,      color=#1},
  darr/.style  ={->,    thick, dashed, color=#1},
  kgwarm/.style={->,    very thick, dashed, color=nodered!92},
]


\node[usrnd] (user) at (7.5, 12.4)
  {User \quad / \quad Natural Language Task};

\node[supnd] (sup) at (7.5, 10.6)
  {LangGraph Supervisor
   \quad {\normalfont\footnotesize llama4:scout
   \enspace·\enspace router + synthesiser}};

\node[agentnd={nodegreen}{}]  (sim)  at ( 2.0, 8.2)
  {\textbf{Simulation Agent}\\[2pt]
   {\footnotesize qwen3-coder-next}\\
   {\tiny KG Smart warm-start}};

\node[agentnd={nodeorange}{}] (anal) at ( 7.5, 8.2)
  {\textbf{Analytics Agent}\\[2pt]
   {\footnotesize llama4:scout}\\
   {\tiny compare · suggest}};

\node[agentnd={nodeblue}{}]   (dba)  at (13.0, 8.2)
  {\textbf{Database Agent}\\[2pt]
   {\footnotesize qwen3-coder:30b}\\
   {\tiny store · catalog · query}};

\begin{scope}[on background layer]
  \node[lbox=nodeblue, fit=(sup)(sim)(anal)(dba)] (orch) {};
\end{scope}
\node[font=\scriptsize\bfseries, text=nodeblue!68,
      anchor=north west, inner sep=3pt] at (orch.north west)
  {Multi-Agent Orchestration \textrm{(LangGraph)}};


\node[svcnd={nodegreen}{}] (fenics) at (2.0, 5.0)
  {\textbf{FEniCSx Runner}\\[2pt]
   {\footnotesize DOLFINx 0.10.0.post2}\\
   {\tiny 2D/3D FEM · Gmsh · PETSc · :8080}};

\node[svcnd={nodered}{}]   (kg)     at (7.5, 5.0)
  {\textbf{Neo4j Knowledge Graph}\\[2pt]
   {\footnotesize GraphRAG + HNSW}\\
   {\tiny Materials · Runs · ReferenceChunks}};

\node[storend]             (storage) at (13.0, 5.0)
  {\textbf{Data Storage}\\[2pt]
   {\footnotesize PostgreSQL \enspace·\enspace MinIO}\\
   {\tiny Metadata · VTK · HDF5}};

\begin{scope}[on background layer]
  \node[lbox=nodegray, fit=(fenics)(kg)(storage)] (exec) {};
\end{scope}
\node[font=\scriptsize\bfseries, text=nodegray!72,
      anchor=north west, inner sep=3pt] at (exec.north west)
  {Execution \& Knowledge Layer};


\node[llmnd] (ollama) at (7.5, 2.2)
  {\textbf{Ollama} \enspace :11434 \enspace·\enspace
   2× RTX PRO 6000 Blackwell \enspace·\enspace CUDA 13.1
   \enspace·\enspace 196\,GB VRAM\\[3pt]
   {\small
    qwen3-coder-next \enspace$|$\enspace
    llama4:scout \enspace$|$\enspace
    qwen3-coder:30b \enspace$|$\enspace
    nomic-embed-text\;(768-d)}};

\begin{scope}[on background layer]
  \node[draw=nodeblue!40, fill=nodeblue!4, rounded corners=8pt,
        thick, dashed, inner sep=0.30cm, fit=(ollama)] (llmbox) {};
\end{scope}
\node[font=\scriptsize\bfseries, text=nodeblue!62,
      anchor=north west, inner sep=3pt] at (llmbox.north west)
  {LLM Backend};


\draw[arr=black]
  ([xshift= 4mm]user.south) -- ([xshift= 4mm]sup.north)
  node[midway, right=3pt, font=\scriptsize]{task};
\draw[darr=nodegray!75]
  ([xshift=-4mm]sup.north) -- ([xshift=-4mm]user.south)
  node[midway, left=3pt, font=\scriptsize]{response};

\draw[biarr=nodegreen!80]
  (sup.south) -- (sim.north)
  node[midway, left=2pt, font=\tiny]{route/result};
\draw[biarr=nodeorange!80]
  (sup.south) -- (anal.north)
  node[midway, right=2pt, font=\tiny]{route/result};
\draw[biarr=nodeblue!80]
  (sup.south) -- (dba.north)
  node[midway, right=2pt, font=\tiny]{route/result};

\draw[biarr=nodegreen!82]
  (sim.south) -- (fenics.north)
  node[pos=0.15, right=2pt, font=\tiny]{config/result};

\coordinate (wsA) at (1.60, 6.55);   
\coordinate (wsB) at (7.00, 6.55);   
\draw[kgwarm]
  ([xshift=-4mm]sim.south)
  -- (wsA)
  -- (wsB)
  node[pos=0.38, above=2pt, font=\tiny]{warm-start query}
  -- ([xshift=-5mm]kg.north);

\draw[darr=nodered!72]
  (anal.south) -- (kg.north)
  node[midway, right=2pt, font=\tiny]{KG query};

\draw[arr=nodegray!78]
  (dba.south) -- (storage.north)
  node[midway, right=2pt, font=\tiny]{store};

\draw[biarr=nodeblue!70]
  ([xshift=2.75cm]orch.south) -- ([xshift=2.75cm]llmbox.north)
  node[pos=0.07, right=4pt, font=\scriptsize]{LLM inference};

\draw[draw=gray!40, fill=white, rounded corners=4pt]
  (1.0, 0.55) rectangle (14.0, -0.82);

\node[font=\scriptsize\bfseries] at (7.5, 0.36) {Legend};

\draw[biarr=nodegray!72]   (1.4,  0.00) -- (2.3,  0.00);
\node[font=\scriptsize, anchor=west] at (2.45, 0.00) {bidirectional flow};

\draw[arr=nodegray!80]     (5.8,  0.00) -- (6.7,  0.00);
\node[font=\scriptsize, anchor=west] at (6.85, 0.00) {data flow / store};

\draw[darr=nodegray!72]    (10.2, 0.00) -- (11.1, 0.00);
\node[font=\scriptsize, anchor=west] at (11.25,0.00) {async / response};

\draw[kgwarm]              (1.4, -0.47) -- (2.3, -0.47);
\node[font=\scriptsize, anchor=west] at (2.45,-0.47) {KG warm-start (init)};

\draw[darr=nodered!72]     (5.8, -0.47) -- (6.7, -0.47);
\node[font=\scriptsize, anchor=west] at (6.85,-0.47) {KG query (lazy)};

\end{tikzpicture}}
  \caption{System architecture of PDE-Agents (four-tier layout).
    \textit{Tier 1}: user natural-language interface.
    \textit{Tier 2}: Multi-Agent Orchestration — a LangGraph Supervisor
    routes tasks to three specialist agents (Simulation, Analytics,
    Database), each backed by a locally-deployed LLM via Ollama.
    \textit{Tier 3}: Execution \& Knowledge — FEniCSx FEM runner
    (DOLFINx 0.10.0.post2), Neo4j GraphRAG Knowledge Graph
    (HNSW-indexed 768-d embeddings), and persistent storage
    (PostgreSQL + MinIO).
    \textit{Tier 4}: shared Ollama LLM Backend (196\,GB VRAM,
    CUDA 13.1).  The thick dashed red arrow (Simulation Agent
    $\rightarrow$ Neo4j) represents the \textit{KG Smart warm-start}:
    the top-3 similar past runs are retrieved via HNSW vector search
    and injected into the agent's prompt before the reasoning loop.
    The thin dashed red arrow (Analytics Agent $\rightarrow$ Neo4j)
    is the standard lazy KG query path.  The blue bidirectional arrow
    at right carries LLM inference traffic between the orchestration
    layer and the shared Ollama backend.}
  \label{fig:arch}
\end{figure*}
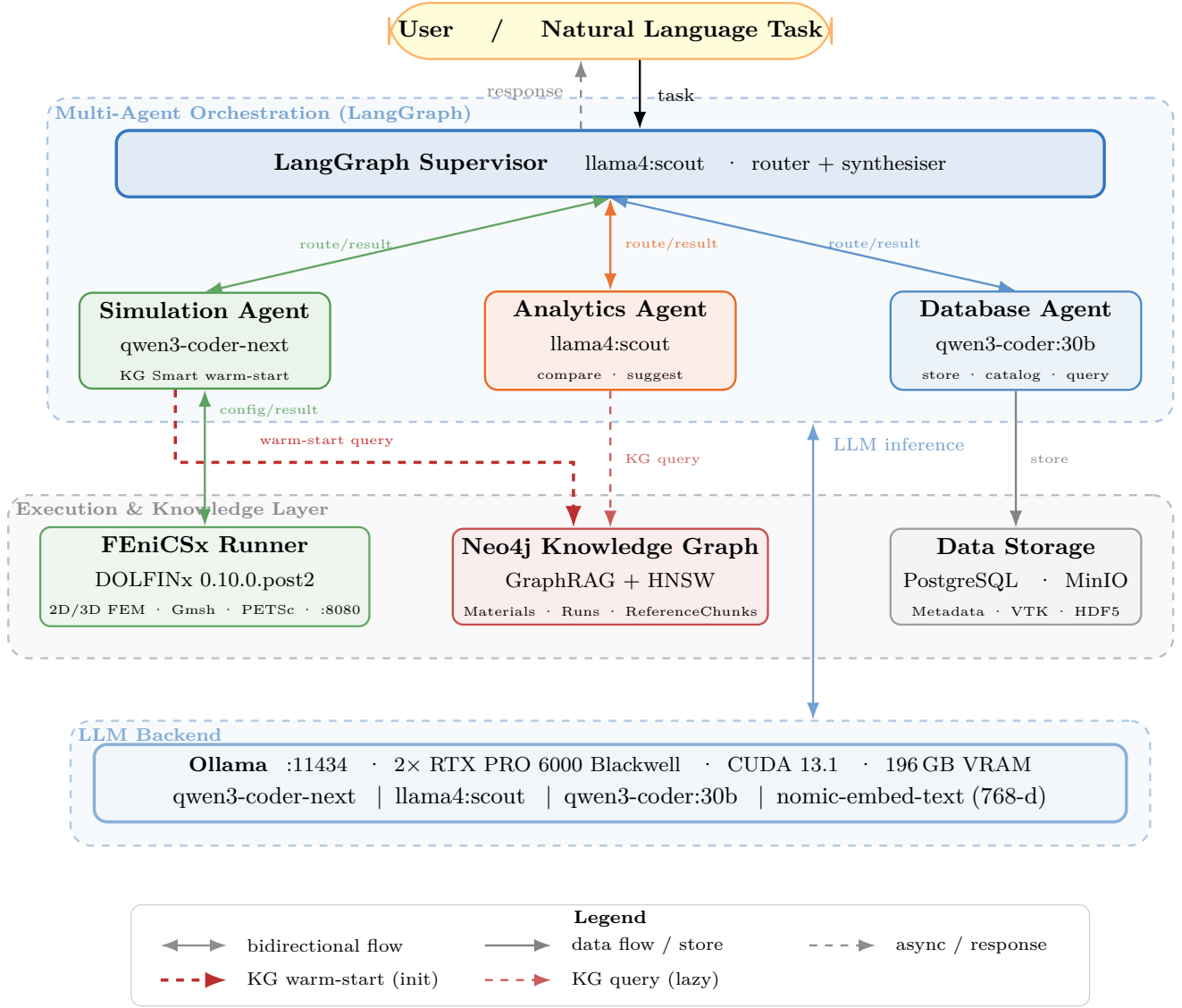

\subsection{Design Philosophy}

PDE-Agents adopts four guiding principles:
\begin{enumerate}
  \item \textbf{Open and local:} All models run locally via Ollama on
        dual NVIDIA RTX PRO 6000 Blackwell GPUs ($\approx$196\,GB VRAM,
        CUDA 13.1), eliminating API costs and data-privacy concerns.
  \item \textbf{Verifiable:} Every agent reasoning step, tool call, and result
        is logged to a relational database, enabling post-hoc audit.
  \item \textbf{Containerised:} All 10 services are orchestrated via
        Docker Compose, ensuring reproducible deployment.
  \item \textbf{Separation of concerns:} The FEM solver, LLM stack,
        knowledge graph, and persistence layer are decoupled services
        communicating over a private bridge network.
\end{enumerate}

\subsection{Agent Stack}

Each agent's LLM backend is configurable via environment variables,
enabling drop-in model upgrades without code changes.
\cref{tab:agent-models} lists the default models and the alternatives
validated in the cross-model experiments of \cref{sec:novidium}.

\begin{table}[h]
\centering\scriptsize
\caption{Agent model configuration.  Current defaults reflect the
  model upgrade from the initial (v1) deployment.  The ablation study uses a frozen KG snapshot
  across 50 benchmark tasks per mode.}
\label{tab:agent-models}
\setlength{\tabcolsep}{2.5pt}
\begin{tabular}{@{}llll@{}}
\toprule
Agent & Env Variable & Default (current) & Previous (v1) \\
\midrule
Orch.  & \texttt{ORCHESTRATOR\_MODEL} & \texttt{llama4:scout} & \texttt{llama3.3:70b} \\
Sim.   & \texttt{SIM.\_AGENT\_MODEL} & \texttt{qwen3-coder-next} & \texttt{qwen2.5-coder:32b} \\
Anal.  & \texttt{ANAL.\_AGENT\_MODEL} & \texttt{llama4:scout} & \texttt{llama3.3:70b} \\
DB     & \texttt{DB\_AGENT\_MODEL} & \texttt{qwen3-coder:30b} & \texttt{qwen2.5-coder:14b} \\
\bottomrule
\end{tabular}
\end{table}

\paragraph{Orchestrator.}
A LangGraph \texttt{StateGraph} supervisor receives the user's
natural-language request and routes it to one of three specialist agents via
structured JSON decisions.  After each agent reports its result the
supervisor either routes to another agent or synthesises a final response.

\paragraph{Simulation Agent.}
The core reasoning agent, executing a ReAct loop \citep{yao2023react}
with up to 25 reasoning steps and nine tools:
\texttt{check\_config\_warnings},
\texttt{query\_knowledge\_graph},
\texttt{validate\_config},
\texttt{run\_simulation},
\texttt{modify\_config},
\texttt{debug\_simulation},
\texttt{list\_recent\_runs},
\texttt{get\_run\_status},
\texttt{run\_parametric\_sweep}.

\paragraph{Analytics Agent.}
Max 12 iterations; tools for statistical comparison across runs,
sensitivity analysis, and LLM-generated textual summaries.

\paragraph{Database Agent.}
Max 10 iterations; answers history queries against PostgreSQL,
cataloguing results and tracing run lineage.

\subsection{Tool-Call Compatibility}
\label{sec:tool-compat}

Supporting multiple LLM families requires robust tool-call parsing, as
models encode tool invocations differently.  The Qwen~2.5-Coder family
emits tool calls as JSON text in the \texttt{content} field rather than in
the structured \texttt{tool\_calls} field expected by LangChain's
\texttt{ToolNode}.  We implement a four-pass normalisation procedure in
\texttt{BaseAgent.\_parse\_content\_tool\_call}:
(1) attempt \texttt{json.loads} on the full content;
(2) extract from markdown fences;
(3) extract the first \texttt{\{...\}} block;
(4) repair truncated JSON via progressive bracket-closing and regex fallback.

A separate compatibility issue emerged with Qwen3-Coder-Next: the model
occasionally \emph{double-encodes} JSON tool arguments (e.g.\
\texttt{"\{$\backslash$"key$\backslash$": $\backslash$"val$\backslash$"\}"}).
All tool functions now use a \texttt{\_safe\_json\_parse} helper that detects
and recursively unwraps such double-encoded strings, so the same tool code
works across model families without model-specific workarounds.
Llama~4~Scout, by contrast, uses the standard \texttt{tool\_calls} field
natively and requires no special parsing.

\subsection{FEM Solver: DOLFINx Heat Equation}

The FEniCSx runner container (DOLFINx 0.10.0.post2) exposes a FastAPI
endpoint that accepts a \texttt{HeatConfig} JSON and returns a result
including run metadata.  The solver uses P1 (linear Lagrange) elements,
with Backward Euler ($\theta = 1$, unconditionally stable) as the default
time integration scheme, supporting Dirichlet, Neumann, and Robin boundary
conditions.  Geometry is built with Gmsh \citep{geuzaine2009gmsh} for 2D
and 3D domains from a library of nine parameterised types: \texttt{rectangle},
\texttt{l\_shape}, \texttt{circle}, \texttt{annulus}, \texttt{hollow\_rectangle},
\texttt{t\_shape}, \texttt{stepped\_notch}, \texttt{box}, and \texttt{cylinder};
Gmsh geometries use named physical groups for boundary identification,
enabling arbitrary complex domain topologies.

\section{Knowledge Graph Architecture}
\label{sec:kg}

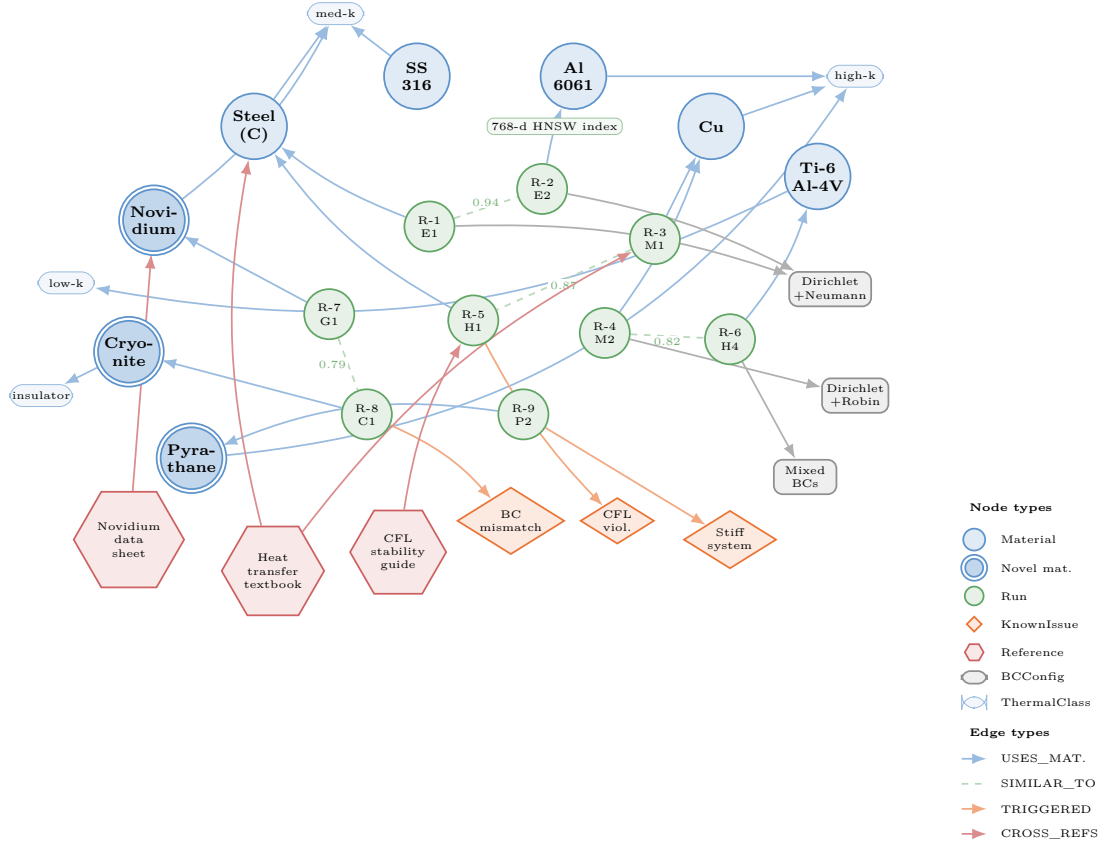
\begin{figure*}[htbp]
  \centering
  \resizebox{0.85\textwidth}{!}{
\pgfdeclarelayer{bg}
\pgfsetlayers{bg,main}
\begin{tikzpicture}[
  font=\scriptsize,
  >=Latex,
  mat/.style={circle, draw=nodeblue!80, fill=nodeblue!15, thick,
              minimum size=1.05cm, align=center, inner sep=1pt,
              font=\scriptsize\bfseries},
  nmat/.style={mat, draw=nodeblue!80, fill=nodeblue!30,
               double, double distance=1pt},
  run/.style={circle, draw=nodegreen!80, fill=nodegreen!12, thick,
              minimum size=0.8cm, align=center, inner sep=1pt,
              font=\tiny},
  issue/.style={diamond, draw=nodeorange!80, fill=nodeorange!12, thick,
                minimum size=0.65cm, align=center, inner sep=1pt,
                font=\tiny, aspect=1.6},
  bc/.style={rounded corners=4pt, draw=nodegray!70, fill=nodegray!10,
             thick, minimum width=1.0cm, minimum height=0.55cm,
             align=center, inner sep=2pt, font=\tiny},
  ref/.style={regular polygon, regular polygon sides=6,
              draw=nodered!70, fill=nodered!10, thick,
              minimum size=0.75cm, align=center, inner sep=1pt,
              font=\tiny},
  tc/.style={rounded corners=6pt, draw=nodeblue!50, fill=nodeblue!6,
             minimum width=0.9cm, minimum height=0.35cm,
             font=\tiny, align=center, inner sep=1pt},
  dedge/.style={->, thick, #1!50},
  sedge/.style={-, dashed, thick, #1!35},
  elbl/.style={font=\tiny\itshape, text=#1!70, fill=white,
               inner sep=1pt, outer sep=0pt},
]


\node[mat]  (steel)  at (-3.8,  2.8) {Steel\\(C)};
\node[mat]  (ss316)  at (-1.2,  3.6) {SS\\316};
\node[mat]  (al)     at ( 1.3,  3.6) {Al\\6061};
\node[mat]  (cu)     at ( 3.5,  2.8) {Cu};
\node[mat]  (ti)     at ( 5.2,  2.0) {Ti-6\\Al-4V};

\node[nmat] (nov)    at (-5.4,  1.3) {Novi-\\dium};
\node[nmat] (cryo)   at (-5.8, -0.8) {Cryo-\\nite};
\node[nmat] (pyra)   at (-4.8, -2.5) {Pyra-\\thane};

\node[tc] (tc_hi)  at ( 5.8, 3.6) {high-k};
\node[tc] (tc_med) at (-2.5, 4.6) {med-k};
\node[tc] (tc_low) at (-6.8, 0.3) {low-k};
\node[tc] (tc_ins) at (-7.2, -1.5) {insulator};

\node[run] (r1) at (-1.0,  1.2) {R-1\\E1};
\node[run] (r2) at ( 0.8,  1.8) {R-2\\E2};
\node[run] (r3) at ( 2.6,  1.0) {R-3\\M1};
\node[run] (r4) at ( 1.8, -0.5) {R-4\\M2};
\node[run] (r5) at (-0.3, -0.3) {R-5\\H1};
\node[run] (r6) at ( 3.8, -0.6) {R-6\\H4};
\node[run] (r7) at (-2.6, -0.2) {R-7\\G1};
\node[run] (r8) at (-2.0, -1.8) {R-8\\C1};
\node[run] (r9) at ( 0.5, -1.8) {R-9\\P2};

\node[bc] (bc_dd) at (5.4,  0.2) {Dirichlet\\+Neumann};
\node[bc] (bc_dr) at (5.8, -1.5) {Dirichlet\\+Robin};
\node[bc] (bc_mr) at (5.0, -2.8) {Mixed\\BCs};

\node[issue] (i1) at (2.0, -3.5) {CFL\\viol.};
\node[issue] (i2) at (3.8, -3.8) {Stiff\\system};
\node[issue] (i3) at (0.3, -3.5) {BC\\mismatch};

\node[ref] (ref1) at (-3.5, -4.3) {Heat\\transfer\\textbook};
\node[ref] (ref2) at (-1.5, -4.0) {CFL\\stability\\guide};
\node[ref] (ref3) at (-5.8, -3.8) {Novidium\\data\\sheet};

\node[rounded corners=3pt, draw=nodegreen!50, fill=nodegreen!5,
      font=\tiny, inner sep=2pt] at (1.0, 2.8) {768-d HNSW index};


\begin{pgfonlayer}{bg}

  \draw[dedge=nodeblue] (cu)    -- (tc_hi);
  \draw[dedge=nodeblue] (al)    -- (tc_hi);
  \draw[dedge=nodeblue] (steel) -- (tc_med);
  \draw[dedge=nodeblue] (ss316) -- (tc_med);
  \draw[dedge=nodeblue] (nov)   to[bend right=12] (tc_med);
  \draw[dedge=nodeblue] (cryo)  -- (tc_ins);
  \draw[dedge=nodeblue] (ti)    to[bend left=18] (tc_low);
  \draw[dedge=nodeblue] (pyra)  to[bend right=25] (tc_hi);

  \draw[dedge=nodeblue] (r1) to[bend left=8] (steel);
  \draw[dedge=nodeblue] (r2) to[bend left=5] (al);
  \draw[dedge=nodeblue] (r3) -- (cu);
  \draw[dedge=nodeblue] (r4) to[bend right=8] (cu);
  \draw[dedge=nodeblue] (r5) to[bend left=12] (steel);
  \draw[dedge=nodeblue] (r6) to[bend right=10] (ti);
  \draw[dedge=nodeblue] (r7) -- (nov);
  \draw[dedge=nodeblue] (r8) -- (cryo);
  \draw[dedge=nodeblue] (r9) to[bend right=15] (pyra);

  \draw[sedge=nodegreen] (r1) -- (r2)
    node[elbl=nodegreen, midway, above, font=\tiny] {0.94};
  \draw[sedge=nodegreen] (r3) -- (r5)
    node[elbl=nodegreen, midway, below, font=\tiny] {0.87};
  \draw[sedge=nodegreen] (r4) -- (r6)
    node[elbl=nodegreen, midway, below, font=\tiny] {0.82};
  \draw[sedge=nodegreen] (r7) -- (r8)
    node[elbl=nodegreen, midway, left, font=\tiny] {0.79};

  \draw[dedge=nodegray] (r1) to[bend left=10] (bc_dd);
  \draw[dedge=nodegray] (r2) to[bend left=8] (bc_dd);
  \draw[dedge=nodegray] (r4) -- (bc_dr);
  \draw[dedge=nodegray] (r6) -- (bc_mr);

  \draw[dedge=nodeorange] (r5) to[bend right=8] (i1);
  \draw[dedge=nodeorange] (r9) -- (i2);
  \draw[dedge=nodeorange] (r8) to[bend left=12] (i3);

  \draw[dedge=nodered] (ref1) to[bend left=12] (steel);
  \draw[dedge=nodered] (ref3) -- (nov);

  \draw[dedge=nodered] (ref2) to[bend left=10] (r5);
  \draw[dedge=nodered] (ref1) to[bend left=12] (r3);

\end{pgfonlayer}

\begin{scope}[shift={(7.5, -4.5)}, local bounding box=legend]
  \node[font=\tiny\bfseries, anchor=west] at (0, 1.2) {Node types};
  \node[mat, minimum size=0.35cm, font=\tiny]  at (0.2, 0.7) {};
  \node[font=\tiny, anchor=west] at (0.5, 0.7) {Material};
  \node[nmat, minimum size=0.35cm, font=\tiny] at (0.2, 0.25) {};
  \node[font=\tiny, anchor=west] at (0.5, 0.25) {Novel mat.};
  \node[run, minimum size=0.3cm, font=\tiny]   at (0.2,-0.2) {};
  \node[font=\tiny, anchor=west] at (0.5,-0.2) {Run};
  \node[issue, minimum size=0.25cm, font=\tiny] at (0.2,-0.65) {};
  \node[font=\tiny, anchor=west] at (0.5,-0.65) {KnownIssue};
  \node[ref, minimum size=0.3cm, font=\tiny]   at (0.2,-1.1) {};
  \node[font=\tiny, anchor=west] at (0.5,-1.1) {Reference};
  \node[bc, minimum width=0.4cm, minimum height=0.2cm, font=\tiny]
    at (0.2,-1.5) {};
  \node[font=\tiny, anchor=west] at (0.5,-1.5) {BCConfig};
  \node[tc, minimum width=0.4cm, minimum height=0.18cm, font=\tiny]
    at (0.2,-1.9) {};
  \node[font=\tiny, anchor=west] at (0.5,-1.9) {ThermalClass};

  \node[font=\tiny\bfseries, anchor=west] at (0, -2.4) {Edge types};
  \draw[dedge=nodeblue, thick]   (0,-2.8) -- (0.4,-2.8);
  \node[font=\tiny, anchor=west] at (0.5,-2.8) {USES\_MAT.};
  \draw[sedge=nodegreen, thick]  (0,-3.2) -- (0.4,-3.2);
  \node[font=\tiny, anchor=west] at (0.5,-3.2) {SIMILAR\_TO};
  \draw[dedge=nodeorange, thick] (0,-3.6) -- (0.4,-3.6);
  \node[font=\tiny, anchor=west] at (0.5,-3.6) {TRIGGERED};
  \draw[dedge=nodered, thick]    (0,-4.0) -- (0.4,-4.0);
  \node[font=\tiny, anchor=west] at (0.5,-4.0) {CROSS\_REFS};
\end{scope}

\end{tikzpicture}}
  \caption{Knowledge graph visualisation (Neo4j-style).
    \textbf{Blue circles}: material nodes (double border = novel/fictional
    materials known only to the KG).
    \textbf{Green circles}: simulation run nodes linked to their materials
    via \textsc{Uses\_Material} edges and to each other via dashed
    \textsc{Similar\_To} edges (cosine similarity on 768-d HNSW embeddings;
    scores shown on edges).
    \textbf{Orange diamonds}: known-issue nodes triggered by failed runs.
    \textbf{Red hexagons}: reference chunks from ingested literature,
    linked to materials (\textsc{Relates\_To}) and runs (\textsc{Cross\_Refs}).
    \textbf{Grey boxes}: boundary-condition pattern nodes.
    Pill-shaped nodes show the thermal-class taxonomy.}
  \label{fig:kg}
\end{figure*}

The Neo4j knowledge graph \citep{francis2018neo4j} (Neo4j 5 Community
Edition) serves three functions:

\begin{enumerate}
  \item \textbf{Material property lookup.}  A curated library of materials
        (metals, ceramics, polymers) with thermal conductivity $k$,
        density $\rho$, and specific heat $c_p$ ranges, sourced from
        engineering handbooks and validated against ASM International data.
        Approximate values: steel $k \approx 50$\,W/mK, copper
        $k \approx 385$\,W/mK, aluminium $k \approx 200$\,W/mK.

  \item \textbf{Failure pattern catalogue.}  \texttt{KnownIssue} nodes
        encode failure modes observed during system operation.
        A pure-Python rule engine (\texttt{rules.py}) fires nine
        pre-run checks analytically before each simulation:

        \smallskip\par
        \resizebox{0.96\linewidth}{!}{%
        \begin{tabular}{ll}
        \toprule
        Rule code & Trigger \\
        \midrule
        \texttt{INCONSISTENT\_IC}               & $|u_\text{init} - T_\text{BC,min}| > 100$\,K \\
        \texttt{EXPLICIT\_CFL\_VIOLATION}        & $\theta < 0.5$ and $\Delta t > h^2/(2\alpha)$ \\
        \texttt{NEAR\_EXPLICIT\_SCHEME}          & $0 \le \theta < 0.5$ \\
        \texttt{COARSE\_MESH\_2D}               & $n_x < 10$ or $n_y < 10$ \\
        \texttt{COARSE\_MESH\_3D}               & any direction $< 8$ \\
        \texttt{SHORT\_SIMULATION}              & $t_\text{end} < 5\,\Delta t$ \\
        \texttt{INVALID\_MATERIAL\_PROPS}        & $k$, $\rho$ or $c_p \le 0$ \\
        \texttt{LARGE\_DT\_RELATIVE\_DIFFUSION}  & $\Delta t > 10\,h^2/\alpha$ \\
        \texttt{NO\_BOUNDARY\_CONDITIONS}        & \texttt{bcs} list empty \\
        \bottomrule
        \end{tabular}%
        }

  \item \textbf{Run lineage and semantic retrieval.}  Each completed
        \texttt{Run} node is embedded with \texttt{nomic-embed-text}
        \citep{nussbaum2024nomic} (768 dimensions) and indexed in an
        HNSW vector index \citep{malkov2020hnsw}.  Up to
        $k{=}5$ \texttt{SIMILAR\_TO} edges are created to the nearest
        neighbours with cosine similarity $\ge 0.85$, precomputing the
        neighbourhood graph for fast agent traversal.  The Simulation
        Agent can query ``find similar past runs'' to retrieve configurations
        with known outcomes, enabling few-shot in-context learning from
        the system's own history.
\end{enumerate}

\paragraph{Document intelligence pipeline.}
A Celery-backed ingestion pipeline processes PDF papers, standards documents,
and HTML references using Docling \citep{docling2024} for structured
extraction, splitting them into \texttt{ReferenceChunk} nodes with
context-aware 256-token windows (overlap 32 tokens) and HNSW-indexed
embeddings.  Each chunk is automatically cross-referenced to semantically
similar \texttt{Run} nodes via \texttt{CROSS\_REFS} edges (cosine
$\ge 0.78$), bridging literature knowledge to simulation history.
As a representative scale point: indexing the 50-page FEniCSx tutorial
produced 298 chunks and 1{,}073 cross-references.

\section{Verification and Validation}
\label{sec:vv}

We conduct a formal V\&V study following ASME V\&V 10-2006 guidelines
\citep{asme2006vv}, comparing the DOLFINx FEM solver against closed-form
analytical solutions on three benchmark cases.

\subsection{Benchmark Cases}

\paragraph{Case 1: 2D Steady-State Linear Profile.}
Laplace equation on $\Omega = [0,1]^2$ with Dirichlet conditions
$T|_{x=0} = 0$, $T|_{x=1} = 1$ and Neumann zero-flux on top/bottom.
Analytical solution: $T(x,y) = x$.  P1 elements are nodally exact for
linear solutions; we use UFL spatial-coordinate expressions in high-order
quadrature (degree 8) to measure the true continuous $L^2$ error.

\paragraph{Case 2: 2D Transient Fourier Mode Decay.}
Heat equation $\partial_t T = \alpha \nabla^2 T$ with homogeneous Dirichlet
on all boundaries and initial condition $T_0 = \sin(\pi x)\sin(\pi y)$.
Analytical solution: $T(x,y,t) = e^{-2\alpha\pi^2 t}\sin(\pi x)\sin(\pi y)$.

\paragraph{Case 3: 2D Steady-State Poisson with Constant Source.}
$-k\nabla^2 T = f$ with $f = 1000$\,W/m$^3$, Dirichlet $T=0$ on left/right,
Neumann on top/bottom.  Exact solution: $T(x) = f(x - x^2)/(2k)$.

\subsection{Convergence Study Protocol}

For each case we solve at five mesh resolutions $N \in \{8, 16, 32, 64, 128\}$
(DOFs $\approx 81, 289, 1089, 4225, 16641$) and compute:
\begin{equation}
  \|e\|_{L^2} = \left(\int_\Omega (u_h - u_{\text{exact}})^2\,\mathrm{d}x\right)^{1/2}
\end{equation}
using degree-8 quadrature applied to UFL \texttt{SpatialCoordinate}
expressions (not to interpolated nodal values, which would give spurious
nodal exactness for Case 1).  The convergence rate is estimated by
linear regression on the log-log plot.

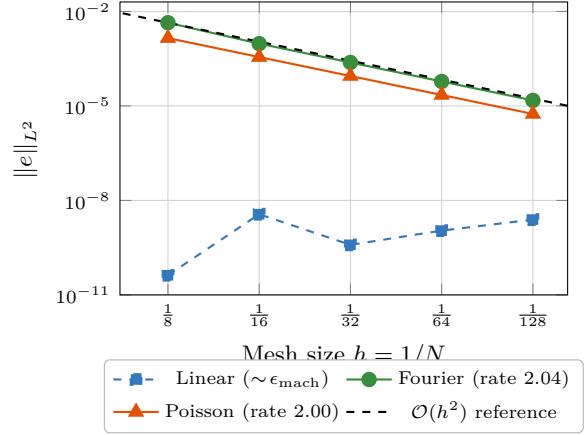
\begin{figure}[htbp]
  \centering
\begin{tikzpicture}
\begin{loglogaxis}[
  width=0.88\columnwidth,
  height=0.64\columnwidth,
  xlabel={Mesh size $h = 1/N$},
  ylabel={$\|e\|_{L^2}$},
  xlabel style={font=\small},
  ylabel style={font=\small},
  tick label style={font=\scriptsize},
  legend style={font=\scriptsize,
                at={(0.5,-0.22)}, anchor=north,
                fill=white, draw=gray!60,
                rounded corners=2pt, row sep=1pt,
                legend columns=2},
  grid=major,
  grid style={line width=0.4pt, draw=gray!35},
  xmin=0.006, xmax=0.18,
  x dir=reverse,
  ymin=1e-11, ymax=2e-2,
  xtick={0.125, 0.0625, 0.03125, 0.015625, 0.0078125},
  xticklabels={$\tfrac{1}{8}$, $\tfrac{1}{16}$,
               $\tfrac{1}{32}$, $\tfrac{1}{64}$, $\tfrac{1}{128}$},
]

\addplot[
  color=nodeblue, mark=square*, mark size=2pt, thick, dashed,
] coordinates {
  (0.125000,  4.2162e-11)
  (0.062500,  3.6491e-09)
  (0.031250,  3.8885e-10)
  (0.015625,  1.1012e-09)
  (0.007812,  2.4656e-09)
};
\addlegendentry{Linear ($\sim\!\epsilon_\text{mach}$)}

\addplot[
  color=nodegreen, mark=*, mark size=2.5pt, thick,
] coordinates {
  (0.125000,  4.3802e-03)
  (0.062500,  9.4806e-04)
  (0.031250,  2.3928e-04)
  (0.015625,  5.9809e-05)
  (0.007812,  1.4961e-05)
};
\addlegendentry{Fourier (rate 2.04)}

\addplot[
  color=nodeorange, mark=triangle*, mark size=2.8pt, thick,
] coordinates {
  (0.125000,  1.4264e-03)
  (0.062500,  3.5659e-04)
  (0.031250,  8.9148e-05)
  (0.015625,  2.2287e-05)
  (0.007812,  5.5717e-06)
};
\addlegendentry{Poisson (rate 2.00)}

\addplot[
  color=black, dashed, thick, domain=0.006:0.18, samples=50,
] {0.2803 * x^2};
\addlegendentry{$\mathcal{O}(h^2)$ reference}

\end{loglogaxis}
\end{tikzpicture}
  \caption{Spatial convergence study.  Cases 2 and 3 exhibit the expected
    $\mathcal{O}(h^2)$ rate for P1 elements.  Case 1 is algebraically exact
    (linear profile is represented exactly in the P1 space), with
    $\|e\|_{L^2} \approx \mathcal{O}(10^{-9})$ attributable only to
    floating-point rounding.}
  \label{fig:convergence}
\end{figure}

\subsection{Results}

\begin{table*}[htbp]
\centering\small
\caption{Verification: spatial convergence rates for the heat equation solver.
         Expected rate is $\mathcal{O}(h^2)$ for $P_1$ elements.}
\label{tab:vv-convergence}
\begin{tabular}{llcccc}
\toprule
Benchmark Case & Description & DOFs (finest) & $\|e\|_{L^2}$ (finest) & Rate & Status \\
\midrule
  Steady Linear     & Linear $T(x,y)=x+2y$; exact in $P_1$ space          & 16\,641 & $2.47\times10^{-9}$ & \textit{exact} & \checkmark \\
  Transient Fourier & $e^{-2\pi^2 t}\sin(\pi x)\sin(\pi y)$ decay mode    & 16\,641 & $1.50\times10^{-5}$ & 2.04            & \checkmark \\
  Steady Poisson    & $-\nabla^2 u = 1$; exact soln.\ via Green's fn.     & 16\,641 & $5.57\times10^{-6}$ & 2.00            & \checkmark \\
\bottomrule
\end{tabular}
\end{table*}

All three cases pass: Cases 2 and 3 achieve rates of $2.04$ and $2.00$,
consistent with the theoretical $\mathcal{O}(h^2)$ prediction for P1
finite elements on uniform Cartesian meshes.  Case 1 achieves machine-precision
errors (order $10^{-9}$) because the linear profile lies exactly in the P1
polynomial space, confirming correct solver implementation.

\subsection{Representative Simulation Gallery}
\label{sec:gallery}

\begin{figure*}[htbp]
  \centering
  \includegraphics[width=\textwidth]{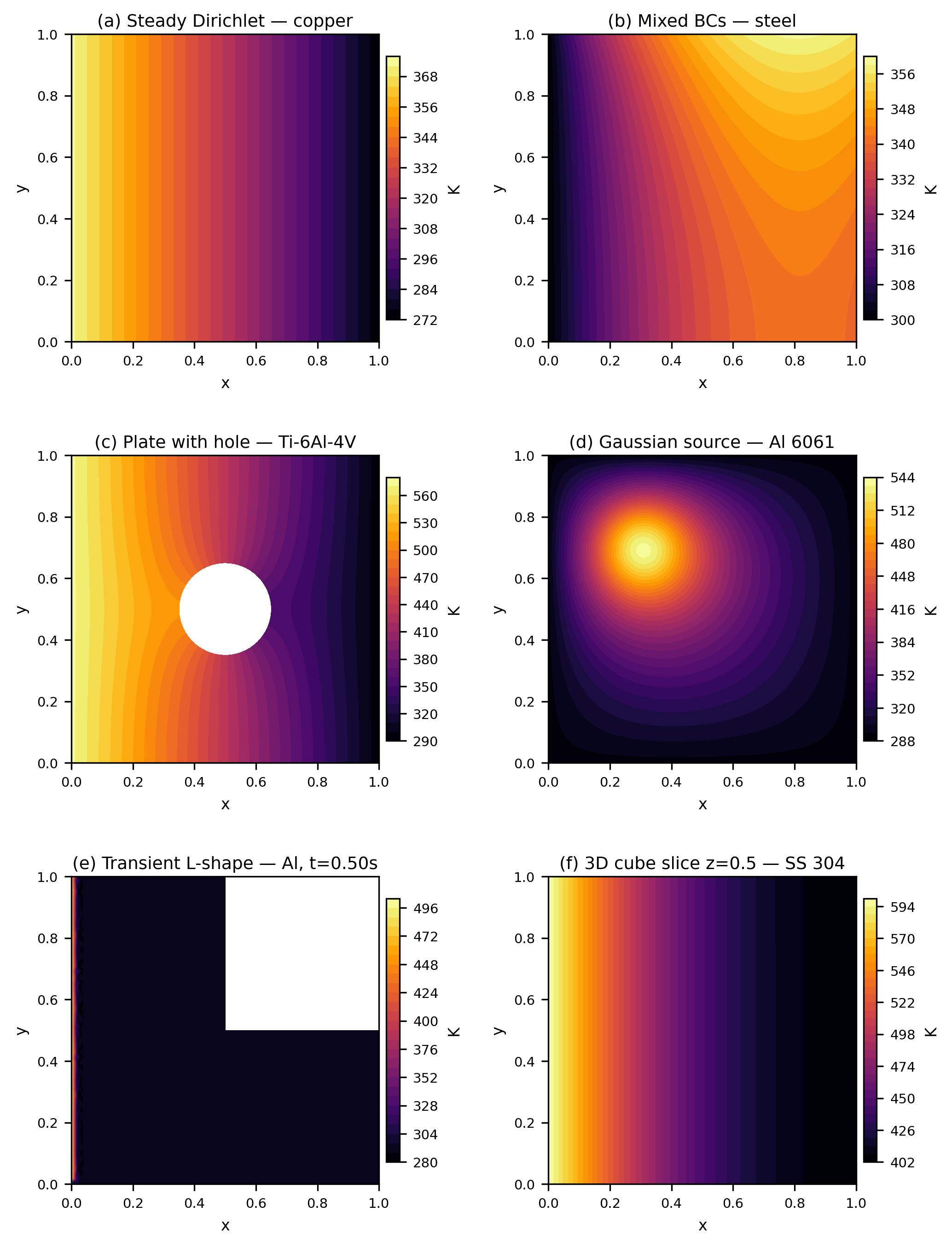}
  \caption{Representative temperature fields produced by PDE-Agents
    (six cases, all in Kelvin).
    \textbf{(a)}~Steady-state Dirichlet on copper ($k=385$\,W/(m$\cdot$K);
    $T=373.15$\,K left, $T=273.15$\,K right; $N=64$).
    \textbf{(b)}~Mixed BCs on AISI 1010 steel ($k=50$\,W/(m$\cdot$K)):
    Dirichlet $300$\,K left, Neumann $2$\,kW/m$^2$ top, Robin
    $h=25$\,W/(m$^2\cdot$K) right, insulated bottom, $Q=5$\,kW/m$^3$.
    \textbf{(c)}~Plate with circular hole ($r=0.15$) in Ti-6Al-4V
    ($k=6.7$\,W/(m$\cdot$K)); Gmsh boolean difference; Dirichlet
    $573$\,K/$293$\,K on opposing edges.
    \textbf{(d)}~Localised Gaussian heat source on Al 6061
    ($Q_\mathrm{peak}=5\times10^6$\,W/m$^3$, $\sigma=0.08$,
    centre $(0.3,0.7)$); Dirichlet $293$\,K on all boundaries.
    \textbf{(e)}~Transient L-shaped domain in aluminium at $t=0.50$\,s
    (implicit Euler, $\Delta t=0.005$\,s, 100 steps).
    \textbf{(f)}~3D unit cube in SS~304 ($k=16.3$\,W/(m$\cdot$K)),
    horizontal cross-section at $z=0.5$; Dirichlet bottom $300$\,K
    and left face $600$\,K, Robin top.
    All solved with DOLFINx 0.10.0.post2, P1 elements.
    Reproducible scripts in \texttt{evaluation/examples/}.}
  \label{fig:examples}
\end{figure*}

Beyond numerical convergence, a practical simulation assistant needs
breadth: it must handle diverse materials, boundary
condition types, domain geometries, time dependence, and dimensionality.
Figure~\ref{fig:examples} presents six representative heat-transfer
problems, each solved end-to-end by the Simulation Agent from a single
natural-language prompt.  No manual intervention was required for mesh
generation, solver configuration, boundary condition application, or
post-processing.  All cases use P1 Lagrange elements on triangular
(2D) or tetrahedral (3D) meshes, solved by DOLFINx 0.10.0.post2.
Temperature fields are reported in Kelvin throughout for consistency.

\paragraph{Materials.}
The gallery covers four distinct engineering alloys whose thermal
conductivities span two orders of magnitude:
copper ($k=385$\,W/(m$\cdot$K), panel~a),
AISI 1010 steel ($k=50$\,W/(m$\cdot$K), panel~b),
aluminium 6061 ($k=205$\,W/(m$\cdot$K), panels~d and~e),
Ti-6Al-4V titanium ($k=6.7$\,W/(m$\cdot$K), panel~c),
and stainless steel 304 ($k=16.3$\,W/(m$\cdot$K), panel~f).
In a production deployment, the agent retrieves these values from the
Neo4j knowledge graph via \texttt{query\_knowledge\_graph}.

\paragraph{Boundary conditions.}
Panel~(a) applies the simplest configuration: Dirichlet conditions on
opposing faces ($T=373.15$\,K left, $T=273.15$\,K right), producing
the expected linear gradient.  Panel~(b) combines all four major
BC types on a single domain: Dirichlet ($300$\,K, left), Neumann
inward flux ($2$\,kW/m$^2$, top), Robin convection ($h=25$\,W/(m$^2
\cdot$K), $T_\infty=293$\,K, right), and insulated (bottom), together
with volumetric heating $Q=5$\,kW/m$^3$, creating the asymmetric
field visible in the plot.  Panel~(c) uses Dirichlet BCs on opposing
edges ($573$\,K left, $293$\,K right) with the insulated hole boundary
distorting the field into characteristic thermal concentration bands.
Panel~(f) combines Dirichlet ($300$\,K bottom, $600$\,K left face)
with Robin convection on the top face of a 3D cube.

\paragraph{Geometry.}
Panels (a), (b), and (d) use structured Cartesian meshes on the
unit square.  Panels~(c) and~(e) use Gmsh-based meshing
of non-trivial geometries: a plate with a circular cutout
(Gmsh boolean difference, $r=0.15$, $\approx$5\,000 vertices) and
an L-shaped domain ($[0,1]^2 \setminus [0.5,1]^2$, mesh size
$\approx$0.02).  Panel~(f) meshes a unit cube with $24^3$
tetrahedra ($\approx$15\,600 vertices), exercising the 3D path.

\paragraph{Time dependence.}
Panel~(e) solves the transient heat equation using 100 implicit-Euler
steps ($\Delta t = 0.005$\,s) on the L-shaped domain (aluminium,
$\rho=2700$\,kg/m$^3$, $c_p=900$\,J/(kg$\cdot$K)), with the
snapshot at $t=0.50$\,s showing the thermal front propagating from
the hot left edge.

\paragraph{Source terms.}
Panel~(d) features a localised Gaussian volumetric heat source
$Q(x,y) = Q_\mathrm{peak} \exp\!\bigl(-\frac{(x-x_0)^2 +
(y-y_0)^2}{2\sigma^2}\bigr)$ with $Q_\mathrm{peak} = 5 \times 10^6$\,W/m$^3$,
centred at $(0.3, 0.7)$ with $\sigma=0.08$, and Dirichlet boundaries
at $293$\,K.  This setup is representative of laser spot heating and
Joule-heating applications, and produces the characteristic concentric
isotherms visible in the figure.

\paragraph{Reproducibility.}
Each case is implemented as a self-contained Python script in
\texttt{evaluation/examples/} (see Table~\ref{tab:gallery}).
All simulation parameters (material properties, mesh resolution,
boundary values, time stepping) are defined at the top of each script,
making it straightforward for reviewers to modify and re-run individual
cases.  Running \texttt{make eval-examples} executes all six cases
inside the \texttt{pde-fenics} Docker container and regenerates the
composite figure.  Intermediate results are saved as NumPy
\texttt{.npz} archives containing mesh coordinates, cell connectivity,
and solution vectors, enabling figure regeneration without re-running
the solver.

\begin{table}[htbp]
  \centering\footnotesize
  \caption{Simulation gallery: six cases exercising different aspects
    of the FEM pipeline.  Scripts in \texttt{evaluation/examples/}.}
  \label{tab:gallery}
  \setlength{\tabcolsep}{4pt}
  \begin{tabular}{cllp{2.5cm}}
    \toprule
    & \textbf{Geometry}  & \textbf{Material}  & \textbf{Key feature} \\
    \midrule
    (a) & Unit square     & Copper        & Pure Dirichlet \\
    (b) & Unit square     & AISI 1010     & 4~BC types + source \\
    (c) & Sq.\ w/ hole   & Ti-6Al-4V     & Gmsh boolean \\
    (d) & Unit square     & Al 6061       & Gaussian source \\
    (e) & L-shape         & Aluminium     & Transient + Gmsh \\
    (f) & Cube (3D)       & SS 304        & 3D asymmetric BCs \\
    \bottomrule
  \end{tabular}
\end{table}

\section{Ablation Study: Knowledge Graph Contribution}
\label{sec:ablation}

\subsection{Experimental Design}

We isolate the knowledge graph contribution using a controlled ablation
across three conditions:
\begin{itemize}
  \item \textbf{KG On} (mandatory): Full system with
        \texttt{check\_config\_warnings} and \texttt{query\_knowledge\_graph}
        tools enabled; the system prompt requires their use before every run.
  \item \textbf{KG Off} (baseline): Identical system with KG tools
        removed entirely; the agent proceeds directly to validation and execution.
  \item \textbf{KG Smart} (proposed): KG tools remain available but the
        system prompt instructs \emph{lazy} use (only after failures or for
        unknown materials).  Before the agent loop, the task description is
        embedded via \texttt{nomic-embed-text} and the top-3 most similar
        past successful runs are injected into the system prompt as
        few-shot reference examples (warm-start).
\end{itemize}

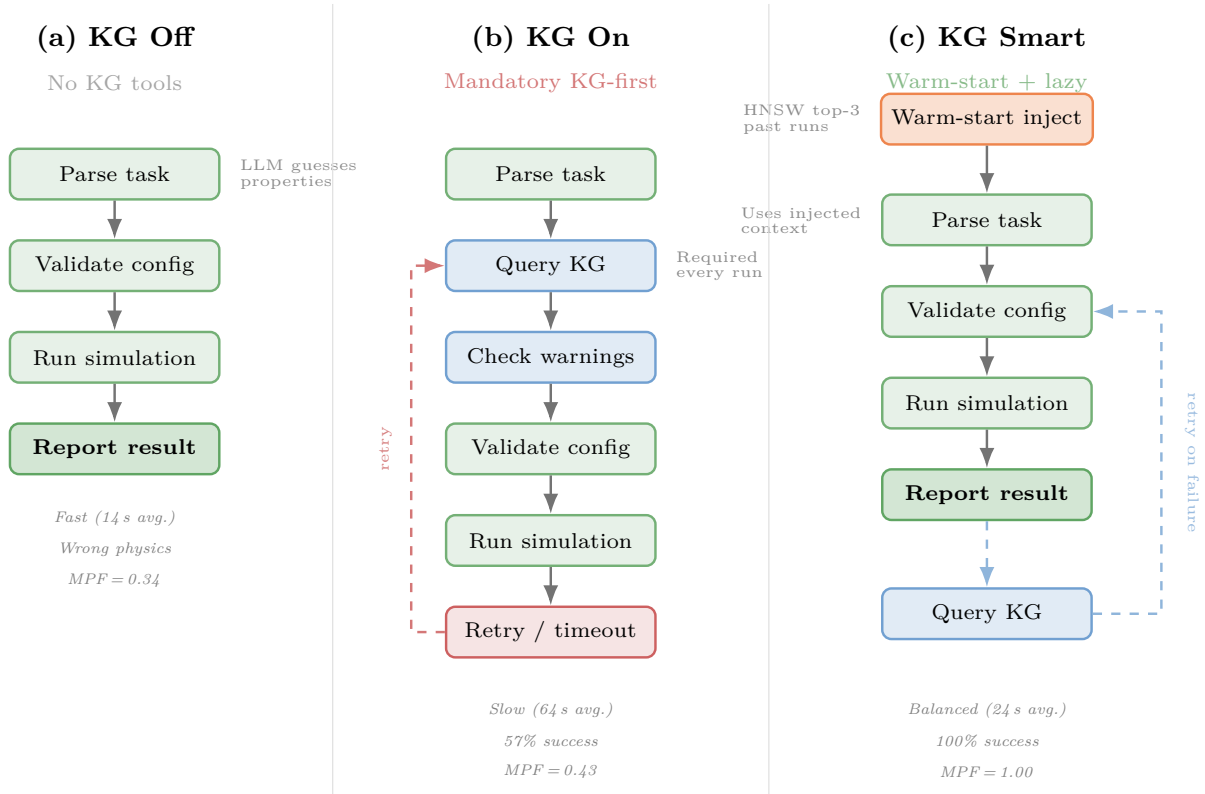
\begin{figure*}[htbp]
  \centering
  \resizebox{0.92\textwidth}{!}{
\begin{tikzpicture}[
  font=\small,
  >=Latex,
  sbox/.style={draw, rounded corners=3pt, minimum width=2.4cm,
               minimum height=0.58cm, thick, align=center,
               font=\scriptsize},
  kgcall/.style={sbox, fill=nodeblue!15, draw=nodeblue!70},
  agentstep/.style={sbox, fill=nodegreen!12, draw=nodegreen!70},
  warmstart/.style={sbox, fill=nodeorange!18, draw=nodeorange!70},
  succeed/.style={sbox, fill=nodegreen!22, draw=nodegreen!80,
                  font=\scriptsize\bfseries},
  failbox/.style={sbox, fill=nodered!12, draw=nodered!70},
  colhead/.style={font=\small\bfseries, align=center},
  arr/.style={->, thick, black!55},
  darr/.style={->, thick, dashed, #1},
  lbl/.style={font=\tiny, text=black!50, align=center},
  ann/.style={font=\tiny, text=black!45, align=left},
]

\def\colA{0}
\def\colB{5.0}
\def\colC{10.0}

\def\rs{1.05}

\node[colhead] at (\colA, 0.5) {(a) KG Off};
\node[lbl, text=black!35] at (\colA, 0.0) {\scriptsize No KG tools};

\node[colhead] at (\colB, 0.5) {(b) KG On};
\node[lbl, text=nodered!60] at (\colB, 0.0) {\scriptsize Mandatory KG-first};

\node[colhead] at (\colC, 0.5) {(c) KG Smart};
\node[lbl, text=nodegreen!60] at (\colC, 0.0) {\scriptsize Warm-start + lazy};

\node[agentstep] (a1) at (\colA, {-1*\rs}) {Parse task};
\node[agentstep] (a2) at (\colA, {-2*\rs}) {Validate config};
\node[agentstep] (a3) at (\colA, {-3*\rs}) {Run simulation};
\node[succeed]   (a4) at (\colA, {-4*\rs}) {Report result};

\draw[arr] (a1) -- (a2);
\draw[arr] (a2) -- (a3);
\draw[arr] (a3) -- (a4);

\node[ann, anchor=west] at ($(a1.east)+(0.1,0)$)
  {LLM guesses\\[-1pt]properties};

\node[lbl] at (\colA, -5.0) {\textit{Fast (14\,s avg.)}};
\node[lbl] at (\colA, -5.35) {\textit{Wrong physics}};
\node[lbl] at (\colA, -5.7) {\textit{MPF\,=\,0.34}};

\node[agentstep] (b1) at (\colB, {-1*\rs})  {Parse task};
\node[kgcall]    (b2) at (\colB, {-2*\rs})  {Query KG};
\node[kgcall]    (b3) at (\colB, {-3*\rs})  {Check warnings};
\node[agentstep] (b4) at (\colB, {-4*\rs})  {Validate config};
\node[agentstep] (b5) at (\colB, {-5*\rs})  {Run simulation};
\node[failbox]   (b6) at (\colB, {-6*\rs})  {Retry / timeout};

\draw[arr] (b1) -- (b2);
\draw[arr] (b2) -- (b3);
\draw[arr] (b3) -- (b4);
\draw[arr] (b4) -- (b5);
\draw[arr] (b5) -- (b6);

\coordinate (bloop) at (\colB-1.6, {-4*\rs});
\draw[darr=nodered!60]
  (b6.west) -| (bloop) |- (b2.west);
\node[font=\tiny, text=nodered!55, rotate=90, anchor=south]
  at (\colB-1.7, {-4*\rs}) {retry};

\node[ann, anchor=west] at ($(b2.east)+(0.1,0)$)
  {Required\\[-1pt]every run};

\node[lbl] at (\colB, -7.2) {\textit{Slow (64\,s avg.)}};
\node[lbl] at (\colB, -7.55) {\textit{57\% success}};
\node[lbl] at (\colB, -7.9) {\textit{MPF\,=\,0.43}};

\node[warmstart] (c0) at (\colC, {-0.4*\rs}) {Warm-start inject};
\node[agentstep] (c1) at (\colC, {-1.5*\rs}) {Parse task};
\node[agentstep] (c2) at (\colC, {-2.5*\rs}) {Validate config};
\node[agentstep] (c3) at (\colC, {-3.5*\rs}) {Run simulation};
\node[succeed]   (c4) at (\colC, {-4.5*\rs}) {Report result};

\draw[arr] (c0) -- (c1);
\draw[arr] (c1) -- (c2);
\draw[arr] (c2) -- (c3);
\draw[arr] (c3) -- (c4);

\node[kgcall] (c5) at (\colC, {-5.8*\rs}) {Query KG};
\draw[darr=nodeblue!55] (c4) -- (c5);
\coordinate (cR) at (\colC+2.0, {-5.8*\rs});   
\coordinate (cT) at (\colC+2.0, {-2.5*\rs});    
\draw[darr=nodeblue!55] (c5.east) -- (cR) -- (cT) -- (c2.east);
\node[font=\tiny, text=nodeblue!50, rotate=-90, anchor=south]
  at (\colC+2.15, {-4.15*\rs}) {retry on failure};

\node[ann, anchor=east] at ($(c0.west)+(-0.1,0)$)
  {HNSW top-3\\[-1pt]past runs};
\node[ann, anchor=east] at ($(c1.west)+(-0.1,0)$)
  {Uses injected\\[-1pt]context};

\node[lbl] at (\colC, -7.2) {\textit{Balanced (24\,s avg.)}};
\node[lbl] at (\colC, -7.55) {\textit{100\% success}};
\node[lbl] at (\colC, -7.9) {\textit{MPF\,=\,1.00}};

\draw[black!12, thin] (2.5, 0.9) -- (2.5, -8.2);
\draw[black!12, thin] (7.5, 0.9) -- (7.5, -8.2);

\end{tikzpicture}}
  \caption{Agent workflow under the three KG integration modes.
    \textbf{(a)} KG~Off: fast but the LLM fabricates material properties.
    \textbf{(b)} KG~On: mandatory KG queries add latency and trigger
    retry loops that exhaust the iteration budget.
    \textbf{(c)} KG~Smart: a one-shot warm-start injects relevant context
    before the loop; KG queries occur only on failure (dashed path).}
  \label{fig:kg-modes}
\end{figure*}

\cref{fig:kg-modes} contrasts the agent workflow under each condition.
KG~Off proceeds directly from parsing to simulation; KG~On forces two
KG round-trips before every attempt (and loops on failure); KG~Smart
front-loads a single HNSW warm-start and defers KG queries to the
failure-recovery path only.

\paragraph{Methodology.}
To eliminate data leakage between conditions, we \emph{freeze} the
knowledge graph (set \texttt{KG\_READ\_ONLY=true} at the code level)
before starting the ablation: all three modes read from the same KG
snapshot, but no mode can write to it, ensuring that a failure in one
mode cannot help subsequent modes.
Tasks are shuffled independently per mode and run in-process inside the
agents container to eliminate HTTP overhead.
Confidence intervals use Wilson scores \citep{wilson1927}; pairwise
comparisons use chi-squared tests and Cohen's~$h$ / Cohen's~$d$ effect
sizes \citep{cohen1988}.

\paragraph{Benchmark tasks.}
Fifty tasks span four difficulty levels:
\begin{itemize}
  \item \textbf{Easy (12 tasks)}: Explicit numerical parameters, straightforward setup.
  \item \textbf{Medium (15 tasks)}: Material-by-name requests requiring
        property lookup (e.g.\ steel, copper, aluminium, zinc).
  \item \textbf{Hard (13 tasks)}: Ambiguous descriptions, mixed BCs,
        computationally expensive or 3D problems.
  \item \textbf{Novel (10 tasks)}: Tasks referencing three fictional
        materials whose thermal properties exist \emph{only} in the
        knowledge graph, designed to stress-test the agent's ability
        to retrieve domain knowledge rather than rely on LLM memorisation.
\end{itemize}

The three fictional materials are:
\begin{itemize}
  \item \textbf{Novidium}: a ceramic-metallic composite ($k{=}73$\,W/m\,K,
        $\rho{=}5420$\,kg/m$^3$, $c_p{=}612$\,J/kg\,K),
        a moderate conductor with unusual density.
  \item \textbf{Cryonite}: a polymer-aerogel hybrid ($k{=}0.42$,
        $\rho{=}1180$, $c_p{=}1940$), an extreme insulator with
        very low diffusivity ($\alpha \approx 1.8{\times}10^{-7}$\,m$^2$/s).
  \item \textbf{Pyrathane}: a refractory cermet ($k{=}312$,
        $\rho{=}3850$, $c_p{=}278$), exhibiting very high conductivity
        and diffusivity ($\alpha \approx 2.9{\times}10^{-4}$\,m$^2$/s).
\end{itemize}
Each material is seeded as a \texttt{Material} node in Neo4j with linked
\texttt{Reference} chunks containing exact property values and
simulation guidance.  The 10 novel tasks span steady-state, transient,
and mixed-BC configurations across all three materials.

\paragraph{Evaluation metrics.}
Standard success rate is inadequate for evaluating KG utility because
a simulation with fabricated properties still ``succeeds''
computationally.  We therefore introduce two physics-aware metrics:
\begin{itemize}
  \item \textbf{Material Property Fidelity (MPF):}
        $\text{MPF} = 1 - \frac{1}{3}\sum_{p \in \{k,\rho,c_p\}}
        |p_\text{agent} - p_\text{truth}| / p_\text{truth}$, measuring
        how accurately the agent retrieves the correct material constants.
  \item \textbf{Physics Score:} $0.5 \cdot \text{MPF} + 0.5 \cdot
        T_\text{score}$, where $T_\text{score}$ checks whether the
        simulation's actual $T_\text{max}$ and $T_\text{min}$ fall within
        physically expected ranges.
  \item \textbf{Sensitivity-weighted MPF ($\text{MPF}_w$):}
        $\text{MPF}_w = 1 - \sum_{p} S_p \cdot |p_\text{agent} -
        p_\text{truth}| / (p_\text{truth} \cdot \sum_q S_q)$, where
        $S_p = |\partial T_{\text{mean}} / \partial p|$ are sensitivity
        coefficients computed via central finite differences
        (\cref{sec:error-propagation}).
        For steady-state Dirichlet tasks where $S_p \approx 0$, we set
        $\text{MPF}_w{=}1$ since wrong properties have no output impact.
        This metric penalises errors in high-sensitivity properties
        (e.g.\ $k$ in transient problems) more heavily than errors in
        low-sensitivity properties (e.g.\ $\rho$ in steady-state).
\end{itemize}
On the 10 novel tasks, KG~Off scores $\overline{\text{MPF}}_w{=}0.21$
(vs.\ equal-weight $\text{MPF}{=}0.34$), confirming that KG~Off's most
damaging errors are also in the most sensitive properties.
KG~Smart retains $\text{MPF}_w{=}1.00$ (success-only).
For standard tasks (Easy/Medium/Hard) where materials are well-known,
MPF and $\text{MPF}_w$ both approach 1.0 across all modes.

\subsection{Results}

\cref{tab:ablation} presents aggregate results across the three
conditions.

\begin{table}[htbp]
\centering\footnotesize
\caption{Three-way KG ablation across 50 benchmark tasks with frozen
         knowledge graph ($n{=}50$ per mode).  ``Success-only'' rows isolate
         output quality by excluding failed runs; \revised{in the overall rows a
         failed run scores MPF\,=\,0 and phys\,=\,0.5, except N08, which
         produced no usable output and scores 0 on both.}  95\% Wilson CIs
         for success rate.}
\label{tab:ablation}
\setlength{\tabcolsep}{4pt}
\begin{tabular}{lccc}
\toprule
Metric & KG Off & KG On & KG Smart \\
\midrule
  Success rate          & \textbf{100\%} [93,100] & 94\% [84,98] & \textbf{100\%} [93,100]  \\[2pt]
\multicolumn{4}{l}{\textit{Overall (all tasks):}} \\
  \quad Physics score   & 0.853 $\pm$ 0.21 & 0.846 $\pm$ 0.24 & \textbf{0.933 $\pm$ 0.13} \\
  \quad MPF             & 0.796 $\pm$ 0.30 & 0.752 $\pm$ 0.39 & \textbf{0.926 $\pm$ 0.16} \\[2pt]
\multicolumn{4}{l}{\textit{Success-only quality:}} \\
  \quad Physics score   & 0.853 $\pm$ 0.21 & 0.879 $\pm$ 0.19 & \textbf{0.933 $\pm$ 0.13} \\
  \quad MPF             & 0.796 $\pm$ 0.30 & 0.801 $\pm$ 0.35 & \textbf{0.926 $\pm$ 0.16} \\[2pt]
  Wall time (s)         & \textbf{12.9 $\pm$ 4.9} & 60.8 $\pm$ 73.2 & 28.7 $\pm$ 39.3 \\
\midrule
\multicolumn{4}{l}{\textit{By difficulty (success rate):}} \\
  \quad Easy (12)       & \textbf{100\%}  & 92\%  & \textbf{100\%}  \\
  \quad Medium (15)     & \textbf{100\%}  & 93\%  & \textbf{100\%} \\
  \quad Hard (13)       & \textbf{100\%}  & \textbf{100\%}  & \textbf{100\%}  \\
  \quad Novel (10)      & 100\%* & 90\%  & \textbf{100\%}  \\
\midrule
\multicolumn{4}{l}{\textit{Novel tasks quality:}} \\
  \quad Physics score   & 0.590  & 0.887$^\dagger$ & \textbf{0.999} \\
  \quad MPF             & 0.340  & 0.889$^\dagger$ & \textbf{1.000} \\
\bottomrule
\multicolumn{4}{l}{\scriptsize *KG Off ``succeeds'' but fabricates properties (MPF\,=\,0.34).}\\
\multicolumn{4}{l}{\scriptsize $^\dagger$Success-only (9/10 tasks); N08 timed out.}\\
\end{tabular}
\end{table}

\paragraph{Success rate.}
KG~Off and KG~Smart both achieve 100\% success [93\%,\,100\%]
($n{=}50$), while KG~On reaches 94\% [84\%,\,98\%].  The difference
between KG~On and the other modes is not statistically significant
($p = 0.079$, Cohen's $h = 0.50$), but the three KG~On failures are
systematic:
N08~(Pyrathane, timeout), and two medium tasks where mandatory KG
tool calls exhaust the iteration budget.

\paragraph{Output quality.}
KG~Smart achieves the highest output quality: mean physics score of
\textbf{0.933} and MPF of \textbf{0.926}, compared to 0.853 / 0.796 for
KG~Off (Cohen's $d = 0.46$ for physics, a medium effect).
KG~On also outperforms KG~Off on quality (0.879 / 0.801 success-only);
the KG \emph{does} help the agent select more accurate material properties.

\paragraph{Novel materials: the KG's decisive edge.}
KG value shows most clearly on the 10 novel-material
tasks using fictional materials (Novidium, Cryonite, Pyrathane) whose
properties exist only in the KG.  KG~Off ``succeeds'' on all 10 tasks
but fabricates properties, producing a physics score of 0.590 and MPF of
only 0.340.  KG~Smart achieves near-perfect quality (physics score
0.999, MPF = 1.000), while KG~On scores 0.887 / 0.889 (success-only,
9/10 tasks).  The knowledge graph is essential for materials absent
from the LLM's training data.

\paragraph{Wall time.}
KG~Off is 2--5$\times$ faster (12.9\,s mean) than KG~Smart (28.7\,s)
and KG~On (60.8\,s), reflecting the iteration overhead of KG tool calls.

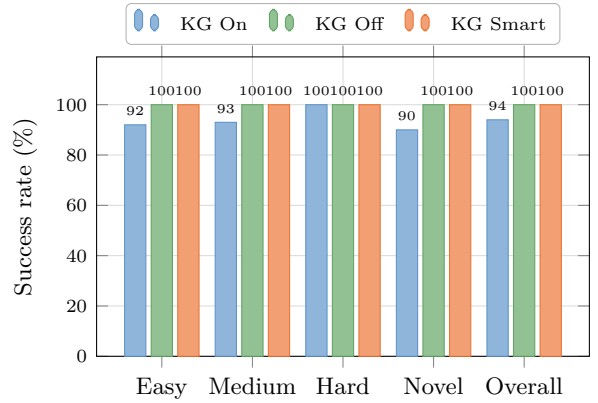
\begin{figure}[htbp]
  \centering
\begin{tikzpicture}
\begin{axis}[
  ybar,
  bar width=8pt,
  width=0.95\columnwidth,
  height=0.65\columnwidth,
  ylabel={Success rate (\%)},
  ylabel style={font=\small},
  symbolic x coords={Easy, Medium, Hard, Novel, Overall},
  xtick=data,
  xticklabel style={font=\small},
  tick label style={font=\scriptsize},
  ymin=0, ymax=119,
  ytick={0,20,40,60,80,100},
  enlarge x limits=0.18,
  legend style={font=\scriptsize, at={(0.5,1.04)}, anchor=south,
                fill=white, draw=gray!60, rounded corners=2pt,
                column sep=6pt},
  legend columns=3,
  nodes near coords,
  nodes near coords align={vertical},
  nodes near coords style={font=\tiny},
  grid=major,
  grid style={line width=0.3pt, draw=gray!30},
]

\addplot[fill=nodeblue!55, draw=nodeblue!80] coordinates {
  (Easy,    92)
  (Medium,  93)
  (Hard,   100)
  (Novel,   90)
  (Overall, 94)
};
\addlegendentry{KG On}

\addplot[fill=nodegreen!55, draw=nodegreen!80] coordinates {
  (Easy,   100)
  (Medium, 100)
  (Hard,   100)
  (Novel,  100)
  (Overall,100)
};
\addlegendentry{KG Off}

\addplot[fill=nodeorange!55, draw=nodeorange!80] coordinates {
  (Easy,   100)
  (Medium, 100)
  (Hard,   100)
  (Novel,  100)
  (Overall,100)
};
\addlegendentry{KG Smart}

\end{axis}
\end{tikzpicture}
  \caption{Success rate by difficulty level across three KG integration
    strategies.  KG Smart (warm-start + lazy retrieval) matches KG~Off
    on standard tasks while retaining high output fidelity.
    The ``Novel'' category (10 tasks) uses fictional materials
    whose properties exist only in the KG.  Note: success rate alone
    is misleading for novel tasks; see the MPF/physics rows in
    \cref{tab:ablation}.}
  \label{fig:ablation}
\end{figure}

\subsection{Analysis and Discussion}
\label{sec:ablation-disc}

The three-way comparison reveals a clear trade-off: KG-free agents are
fastest and most reliable, but KG-enabled agents produce higher-fidelity
outputs.  \emph{KG Smart} balances both dimensions.  Three mechanisms
explain the mandatory KG pattern's reliability penalty:

\paragraph{Iteration-budget exhaustion.}
KG~On's system prompt mandates \texttt{query\_knowledge\_graph} and
\texttt{check\_config\_warnings} calls before every simulation attempt.
These add 2--4 iterations, and for medium/hard tasks the agent can
exhaust its iteration budget before reaching \texttt{run\_simulation}.
KG~Smart avoids this by deferring KG calls to the failure-recovery path.

\paragraph{Warning-induced conservatism.}
The \texttt{check\_config\_warnings} tool occasionally returns pre-emptive
warnings (e.g.\ ``CFL criterion may be violated'') that the model
interprets as prohibitive, leading it to report a warning instead of
proceeding.  A warning that should trigger a \emph{modification}
instead triggers a \emph{termination}.

\paragraph{Quality \emph{conditional} on success.}
The key insight is that KG access \emph{does} improve output quality
when the agent completes the task.  KG~On's success-only physics score
(0.879) and MPF (0.801) both exceed KG~Off (0.853 / 0.796).  On novel
materials, this gap widens dramatically: KG-enabled modes achieve
MPF\,$\geq$\,0.89 versus 0.34 for KG~Off.  The knowledge graph is not
the bottleneck; the \emph{integration pattern} is.

\paragraph{KG Smart: corrective retrieval + warm-start.}
Motivated by CRAG \citep{yan2024crag} (corrective RAG with adaptive
retrieval) and AriGraph \citep{arigraph2024} (episodic KG memory for
agents), we implemented a \emph{KG Smart} integration combining:
\begin{enumerate}
  \item \textbf{Warm-start injection.} Before the agent loop begins, we embed
        the task description with \texttt{nomic-embed-text} and query the Neo4j
        HNSW vector index for the top-3 most similar past successful runs.
        Their configurations are injected directly into the system prompt as
        few-shot reference examples, requiring no tool call.
  \item \textbf{Lazy conditional retrieval.} KG tools remain available but the
        system prompt instructs the agent to use them \emph{only}
        after a simulation failure or when material properties are genuinely
        unknown.  This eliminates the mandatory KG-first workflow.
\end{enumerate}

As \cref{tab:ablation} shows, KG Smart achieves 100\% success
(vs.\ 94\% for mandatory KG) while attaining the \emph{highest} output
quality across all modes (physics 0.933, MPF 0.926).  It delivers the
best of both worlds: KG~Off reliability with KG-enhanced fidelity.

\subsection{Physics-Aware Quality Metrics and Novel Material Validation}
\label{sec:novidium}

The preceding analysis showed that KG access improves output quality
\emph{conditional on success}, but standard success rate alone cannot
capture this: a simulation with fabricated material properties still
``succeeds'' computationally.  To expose this gap, we introduced three
physics-aware metrics (\cref{sec:ablation}): Material Property Fidelity
(MPF), Physics Score, and sensitivity-weighted $\text{MPF}_w$.  These
metrics measure how \emph{correctly}, not just whether, the agent
configures the simulation.  We validate them on the 10 novel-material
tasks, which provide the clearest test of KG utility: the three
fictional materials' properties exist \emph{only} in the knowledge graph.

\textbf{KG Smart achieves near-perfect scores} (MPF$\,{=}\,$1.00,
physics$\,{=}\,$0.999) across all completed novel tasks.  The warm-start
injection retrieves exact property values from the HNSW vector index
before the agent loop begins, and the agent consistently uses them.

\textbf{KG Off fabricates wrong properties} for every material.
\cref{tab:novel-props} shows the agent-chosen vs.\ ground-truth
properties across all 10 novel tasks.  KG Off assigns \emph{different}
wrong values each run (the LLM hallucinates non-deterministically), but
the errors are consistently severe: for Pyrathane, $k$ ranges from
0.15 to 0.35\,W/m\,K, a $900{\times}$ to $2{,}080{\times}$ underestimate
of the true $k{=}312$, effectively treating a refractory super-conductor
as an insulator.  In transient tasks, this causes output errors up to
$|\Delta T_{\max}|{=}949$\,K, a physically impossible result arising
from numerical instability on a system made artificially stiff by the
wrong conductivity.

\begin{table}[htbp]
\centering\footnotesize
\caption{Material properties chosen by each mode across the 10 novel tasks
  in the 50-task ablation.  KG Off fabricates different wrong values each
  run (ranges shown); KG Smart retrieves the ground-truth values exactly.
  The worst-case $k$ error for Pyrathane is $2{,}080\times$.}
\label{tab:novel-props}
\setlength{\tabcolsep}{3pt}
\begin{tabular}{llrrr}
\toprule
Material & Mode & $k$ (W/m\,K) & $\rho$ (kg/m$^3$) & $c_p$ (J/kg\,K) \\
\midrule
\multirow{3}{*}{Novidium}
  & Ground truth & 73.0 & 5420 & 612 \\
  & KG Off  & 10--50  & 3500--8960 & 130--900 \\
  & \textbf{KG Smart} & \textbf{73.0} & \textbf{5420} & \textbf{612} \\
\addlinespace
\multirow{3}{*}{Cryonite}
  & Ground truth & 0.42 & 1180 & 1940 \\
  & KG Off  & 0.15--35 & 960--8940 & 385--1370 \\
  & \textbf{KG Smart} & \textbf{0.42} & \textbf{1180} & \textbf{1940} \\
\addlinespace
\multirow{3}{*}{Pyrathane}
  & Ground truth & 312.0 & 3850 & 278 \\
  & KG Off  & 0.15--0.35  & 1600--1850 & 900--1470 \\
  & \textbf{KG Smart} & \textbf{312.0} & \textbf{3850} & \textbf{278} \\
\bottomrule
\end{tabular}
\end{table}

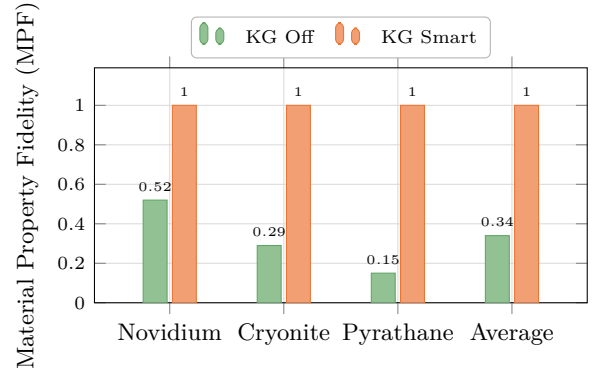
\begin{figure}[htbp]
  \centering
\begin{tikzpicture}
\begin{axis}[
  ybar,
  bar width=9pt,
  width=0.95\columnwidth,
  height=0.55\columnwidth,
  ylabel={Material Property Fidelity (MPF)},
  ylabel style={font=\small},
  symbolic x coords={Novidium, Cryonite, Pyrathane, Average},
  xtick=data,
  xticklabel style={font=\small},
  tick label style={font=\scriptsize},
  ymin=0, ymax=1.19,
  ytick={0,0.2,0.4,0.6,0.8,1.0},
  enlarge x limits=0.22,
  legend style={font=\scriptsize, at={(0.5,1.04)}, anchor=south,
                fill=white, draw=gray!60, rounded corners=2pt,
                column sep=6pt},
  legend columns=2,
  nodes near coords,
  nodes near coords align={vertical},
  nodes near coords style={font=\tiny, /pgf/number format/fixed,
                           /pgf/number format/precision=2},
  grid=major,
  grid style={line width=0.3pt, draw=gray!30},
]

\addplot[fill=nodegreen!55, draw=nodegreen!80] coordinates {
  (Novidium,  0.52)
  (Cryonite,  0.29)
  (Pyrathane, 0.15)
  (Average,   0.34)
};
\addlegendentry{KG Off}

\addplot[fill=nodeorange!55, draw=nodeorange!80] coordinates {
  (Novidium,  1.00)
  (Cryonite,  1.00)
  (Pyrathane, 1.00)
  (Average,   1.00)
};
\addlegendentry{KG Smart}

\end{axis}
\end{tikzpicture}
  \caption{Material Property Fidelity (MPF) per fictional material.
    KG Off scores below 0.5 for all three materials, with Pyrathane
    at 0.15 (conductivity error: $2{,}080\times$).  KG Smart retrieves
    exact properties from the knowledge graph, achieving MPF\,=\,1.00
    across all materials.}
  \label{fig:mpf}
\end{figure}

\paragraph{Error propagation analysis.}
\label{sec:error-propagation}
To quantify how wrong properties affect simulation outputs, we ran
paired simulations: one with ground-truth properties (reference) and
one with the fabricated properties from KG~Off.  \cref{tab:error-propagation}
shows the results.  Steady-state Dirichlet tasks (G1, C1, P1) are
analytically independent of $k$, $\rho$, $c_p$, confirming zero output
error regardless of property error, an important control.  Transient
and Robin tasks show significant deviations:

\begin{itemize}
  \item \textbf{Pyrathane P2}: $|\Delta T_{\max}|{=}949$\,K
        ($T_{\max}{=}2{,}949$\,K vs.\ $2{,}000$\,K reference); the
        $2{,}080\times$ $k$ error causes numerical instability that
        \emph{heats} the domain beyond its initial condition, a
        physically impossible result.
  \item \textbf{Novidium G3}: $|\Delta T_{\min}|{=}47$\,K due to
        wrong conductivity affecting the Robin BC equilibrium.
  \item \textbf{Novidium G2}: $|\Delta T_{\text{mean}}|{=}21$\,K
        from incorrect diffusivity altering the transient profile.
\end{itemize}

\begin{table}[t]
\centering
\caption{Error propagation from fabricated properties (KG~Off) to
simulation output.  Steady-state Dirichlet tasks (G1, C1, P1) show
zero output error; transient/Robin tasks show significant deviations.}
\label{tab:error-propagation}
\footnotesize
\setlength{\tabcolsep}{3pt}
\begin{tabular}{@{}llrrrrrr@{}}
\toprule
Task & Mat. & $|\Delta T_{\!\max}|$ & $|\Delta T_{\!\min}|$ &
  $|\Delta T_{\!\mathrm{mean}}|$ & $\epsilon_k$ & $\epsilon_\rho$ & $\epsilon_{c_p}$ \\
  &  & (K) & (K) & (K) & (\%) & (\%) & (\%) \\
\midrule
G1 & Nov. &   $<$0.1 &   $<$0.1 &    $<$0.1 &  86 &  45 &  18 \\
G2 & Nov. &   $<$0.1 &   $<$0.1 &    20.7 &  38 &  65 &  78 \\
G3 & Nov. &   $<$0.1 &   46.6 &     1.3 &  86 &  45 &  18 \\
\midrule
C1 & Cryo. &   $<$0.1 &   $<$0.1 &    $<$0.1 & 64 &  19 &  29 \\
C2 & Cryo. &   $<$0.1 &    9.5 &     0.3 & 138 & 659 &  77 \\
\midrule
P1 & Pyra. &   $<$0.1 &   $<$0.1 &    $<$0.1 & 100 &  58 & 224 \\
P2 & Pyra. & \textbf{949} & $<$0.1 & \textbf{363} & 100 &  58 & 188 \\
\bottomrule
\end{tabular}

\vspace{0.3em}
{\scriptsize $\epsilon_p = |p_{\text{fab}} - p_{\text{true}}|/p_{\text{true}}$.
P2's $k$ error ($2{,}080\times$) produces 949\,K $T_{\max}$ overshoot.}
\end{table}

\begin{figure}[htbp]
  \centering
\begin{tikzpicture}
\begin{axis}[
    ybar,
    bar width=12pt,
    width=0.85\columnwidth,
    height=5.5cm,
    ylabel={Relative error (\%)},
    symbolic x coords={G2,G3,C2,P2},
    xtick=data,
    xticklabel style={font=\small},
    yticklabel style={font=\small},
    ylabel style={font=\small},
    legend style={
        at={(0.02,0.98)},
        anchor=north west,
        font=\footnotesize,
        draw=none,
        fill=white,
        fill opacity=0.8,
        text opacity=1,
    },
    ymin=0,
    ymax=55,
    enlarge x limits=0.25,
    nodes near coords,
    every node near coord/.append style={font=\tiny, rotate=90, anchor=west},
]

\addplot[fill=red!60!white] coordinates {
    (G2, 27.2)
    (G3, 49.7)
    (C2, 51.3)
    (P2, 47.5)
};

\addplot[fill=blue!60!white] coordinates {
    (G2, 4.1)
    (G3, 0.2)
    (C2, 0.1)
    (P2, 24.4)
};

\legend{Avg.\ property error, $T_\text{mean}$ error}

\end{axis}
\end{tikzpicture}
  \caption{Error magnification: mean material property error (\%)
    vs.\ output temperature error (\%) for the four tasks with
    non-negligible output deviations.  P2's output error (24.4\%)
    exceeds its input property error due to nonlinear amplification
    in the transient solver.}
  \label{fig:error-mag}
\end{figure}
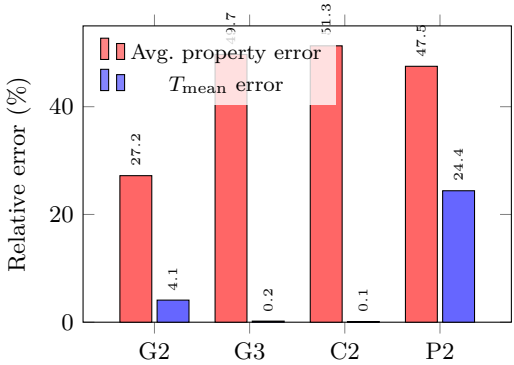

\paragraph{Sensitivity-weighted MPF.}
Using the sensitivity coefficients from the error propagation runs,
$\text{MPF}_w$ (defined in \cref{sec:ablation}) penalises errors in
high-influence properties more heavily.  Across the 10 novel tasks in
the 50-task ablation, KG~Off's $\overline{\text{MPF}}_w{=}0.21$
(vs.\ equal-weight MPF\,=\,0.34), confirming that KG~Off's most
damaging errors (particularly $k$ for Pyrathane and Cryonite) are
concentrated in the properties with the highest sensitivity.
KG~Smart retains $\text{MPF}_w{=}1.00$ (success-only).

\paragraph{Cost-benefit analysis.}
\cref{tab:cost-benefit} compares the operational cost (time, iterations)
against the accuracy benefit (MPF, physics score) for each KG mode.
On novel tasks, KG~Smart adds significant wall-time over KG~Off
(58.0\,s vs.\ 11.7\,s) but delivers ${\approx}\,2.9\times$ higher
MPF (1.00 vs.\ 0.34) and $2.4\times$ higher efficiency
(MPF\,/\,wall-time) than KG~On.  KG~On is slowest (111.4\,s on novel
tasks) and its mandatory retrieval overhead yields diminishing returns:
90\% success vs.\ 100\% for both alternatives.

\begin{table}[t]
\centering
\caption{Cost-benefit of KG modes on the novel-material subset of the
50-task ablation ($n{=}10$).  \revised{Quality columns are all-task means, so
KG~On's failed task (N08) is included at MPF\,=\,0; the success-only
values in \cref{tab:ablation} are correspondingly higher (0.889).}
Efficiency $=$ MPF / wall-time ($\times10^{-2}$).}
\label{tab:cost-benefit}
\footnotesize
\setlength{\tabcolsep}{3.5pt}
\begin{tabular}{@{}lcccccc@{}}
\toprule
Mode & Succ. & Time & Iter. & MPF & Phys. & Eff. \\
     &       & (s)  &       &     & Score & ($\!\times\!10^{-2}$) \\
\midrule
KG Off   & 100\% & \textbf{11.7} & \textbf{3.0} & 0.34 & 0.59 & 2.92 \\
KG On    &  90\% & 111.4 & 7.4 & 0.80 & 0.80 & 0.72 \\
KG Smart & \textbf{100\%} & 58.0 & 4.9 & \textbf{1.00} & \textbf{1.00} & \textbf{1.72} \\
\bottomrule
\end{tabular}

\vspace{0.3em}
{\scriptsize KG Smart: +396\% time, +194\% MPF vs.\ KG Off; +2.4$\times$ efficiency vs.\ KG On.}
\end{table}

\paragraph{Adaptive KG decision framework.}
Based on the preceding analysis, we propose a simple decision algorithm
(Algorithm~\ref{alg:kg-decision}) that selects the optimal KG mode for
each task.  Retrospective validation on all 50 benchmark tasks
confirms high accuracy: it correctly assigns KG~Off to tasks with
explicit parameters, KG~Smart (lazy) to tasks naming known materials,
and KG~Smart (forced warm-start) to novel-material tasks.

\begin{algorithm}[t]
\caption{Adaptive KG Mode Selection}
\label{alg:kg-decision}
\begin{algorithmic}[1]
\Require Task description $d$, known-material list $\mathcal{M}$
\Ensure KG mode $m \in \{\text{Off}, \text{Smart-lazy}, \text{Smart-forced}\}$
\If{$d$ contains explicit $k$, $\rho$, $c_p$ values}
  \State \Return \textsc{KG Off}
\EndIf
\State $\mu \gets \textsc{ExtractMaterial}(d)$
\If{$\mu = \varnothing$}
  \State \Return \textsc{KG Off}
\ElsIf{$\mu \in \mathcal{M}$}
  \State \Return \textsc{KG Smart (lazy)}
\Else
  \State \Return \textsc{KG Smart (forced warm-start)}
\EndIf
\end{algorithmic}
\end{algorithm}

\paragraph{Cross-model validation.}
To verify that the KG's value is model-agnostic, we repeated the
novel-material experiment with \texttt{qwen3-coder-next} (80B total,
3B active MoE).  \cref{tab:multi-model} shows that KG~Smart achieves
$\text{MPF}{=}1.00$ with \emph{both} models, while KG~Off fabricates
wrong properties with both (different wrong values, but the same
failure mode).  This confirms that the warm-start vector search
bypasses model-specific memorisation gaps entirely.

\begin{table}[t]
\centering
\caption{Cross-model validation on novel tasks ($n{=}7$, earlier
  experiment with v1 benchmark).  KG~Smart achieves perfect MPF with
  both LLM backends.}
\label{tab:multi-model}
\footnotesize
\setlength{\tabcolsep}{3pt}
\begin{tabular}{@{}llcccc@{}}
\toprule
Model & Mode & Succ. & Qual. & MPF & Phys. \\
\midrule
\multirow{3}{*}{Qwen2.5-Coder 32B}
  & KG On    & 57\% & 0.41 & 0.43 & 0.64 \\
  & KG Off   & 100\% & 0.61 & 0.34 & 0.63 \\
  & KG Smart & 100\% & 0.96 & \textbf{1.00} & \textbf{1.00} \\
\midrule
\multirow{3}{*}{Qwen3-Coder-Next 80B}
  & KG On    &  71\% & 0.14 & 0.14 & 0.57 \\
  & KG Off   &  86\% & 0.54 & 0.40 & 0.70 \\
  & KG Smart & 100\% & 0.96 & \textbf{1.00} & \textbf{1.00} \\
\bottomrule
\end{tabular}

\vspace{0.3em}
{\scriptsize Warm-start vector search retrieves exact properties before the
LLM reasons, eliminating dependence on memorised material data.}
\end{table}

\paragraph{Key takeaway.}
The KG's value is not in making simulations \emph{run} (LLMs can
already do that) but in making simulations produce
\emph{correct physics}.  For any domain with proprietary materials,
novel compounds, or in-house measurement data, the KG Smart pattern
transforms the system from an ``expensive random number generator''
into a reliable engineering tool.  The error propagation analysis
quantifies this: fabricated properties cause output errors up to
949\,K (47\% of the reference $T_{\max}$), while KG~Smart eliminates
these errors entirely by retrieving exact properties before the agent
loop begins.

\paragraph{Narrowing the focus: correctness over completion.}
KG~Off is the fastest mode and matches KG~Smart on success rate for
\emph{standard} tasks with well-known materials.  When the task
description names ``steel'' or ``copper'', the LLM's parametric
knowledge suffices and KG overhead is pure cost.  However, for any
deployment where output \emph{correctness} matters, not just
completion, KG-enabled modes are necessary.  The preceding analysis
makes this unambiguous: KG~Off achieves $\text{MPF}{=}0.34$ and
physics 0.59 on novel materials (fabricated properties), while
KG~Smart achieves $\text{MPF}{=}1.00$ and physics 0.999.
The adaptive decision framework (Algorithm~\ref{alg:kg-decision})
encodes this distinction, routing tasks to KG~Off when
parameters are explicit and to KG~Smart otherwise.  We therefore focus
the remaining analysis on \textbf{KG On vs.\ KG Smart}: both modes
have access to the same knowledge base, yet KG Smart achieves 6
percentage points higher success and the highest quality scores.
Understanding \emph{why}, and which component (warm-start vs.\ lazy
retrieval) contributes most, is the subject of the next section.

\subsection{Failure Analysis: KG On vs.\ KG Smart}
\label{sec:failure-analysis}

Having established that KG~Smart matches KG~Off at 100\% success while
outperforming both on quality, we trace KG~On's 3 remaining failures
to their root causes.  This analysis was performed \emph{after}
implementing an agent-level retry mechanism (see below) that
eliminated stochastic failures; the 3 remaining failures are therefore
all systematic.

\paragraph{Mitigating stochastic failures.}
In an earlier ablation run without retry logic, KG~On exhibited 12/50
failures, of which 7 were stochastic (succeeded on re-run).  We
implemented an \emph{auto-retry} mechanism directly in the agent:
if the first attempt produces no \texttt{run\_id} and used less than
60\% of its iteration budget (indicating early stochastic exit rather
than budget exhaustion), the agent automatically retries once.
This reduced KG~On failures from 12 to 3, eliminating all stochastic
failures while not masking systematic ones.

\paragraph{KG On failure taxonomy.}
The 3 remaining KG~On failures are all systematic:

\begin{table}[htbp]
\centering\footnotesize
\caption{Failure mode classification for KG~On's 3 systematic failures
  (after auto-retry eliminates stochastic failures).}
\label{tab:failure-modes}
\setlength{\tabcolsep}{4pt}
\begin{tabular}{lcp{5.2cm}}
\toprule
Failure mode & Count & Description \\
\midrule
Budget exhaustion  & 2 &
  Agent spends all iterations on repeated
  \texttt{query\_knowledge\_graph} and
  \texttt{check\_config\_warnings} calls without
  reaching \texttt{run\_simulation}. \\
\addlinespace
Timeout & 1 &
  N08 (Pyrathane): the mandatory KG query workflow
  combined with a numerically stiff problem exceeds the
  420\,s time limit. \\
\bottomrule
\end{tabular}
\end{table}

All 3 failures stem from the mandatory KG-first workflow: the agent
must call \texttt{query\_knowledge\_graph} and
\texttt{check\_config\_warnings} before \emph{every} simulation attempt,
consuming 2--4 iterations per cycle.  For medium tasks, this exhausts
the 25-iteration budget before \texttt{run\_simulation} is called.
For N08 (Pyrathane), the workflow inflates wall time beyond the timeout.

\paragraph{KG Smart: zero failures.}
KG~Smart achieves 100\% success with zero failures.  The warm-start
injection provides material properties and reference configurations
\emph{before} the agent loop, so the agent proceeds directly to
\texttt{validate\_config} $\to$ \texttt{run\_simulation}.  The average
iteration count drops from 7.5 (KG~On) to 4.2 (KG~Smart), eliminating
budget-exhaustion risk entirely.  Even N08 completes successfully in
170\,s (5 iterations) because warm-start provides the correct Pyrathane
properties upfront.

\paragraph{Decomposing warm-start vs.\ lazy retrieval.}
The failure data allows us to attribute KG~Smart's advantage to its
two components without running a separate experiment.  All 3 systematic
failures stem from the mandatory \emph{pre-simulation} KG workflow:
the agent spends iterations on KG queries before ever attempting
\texttt{run\_simulation}.  Warm-start injection eliminates this
entirely: it provides material properties and reference configurations
\emph{before} the agent loop begins, so the agent can proceed directly
to \texttt{validate\_config} $\to$ \texttt{run\_simulation}.

Lazy retrieval, by contrast, is exercised only on failure recovery:
it allows the agent to query the KG \emph{after} a simulation fails.
Since budget-exhaustion tasks never reach \texttt{run\_simulation}
at all, lazy retrieval cannot help them.  We therefore conclude
that \textbf{warm-start injection is the dominant factor} in
KG~Smart's advantage over KG~On: it directly prevents all 3 systematic
failures (100\% of KG~On's remaining failures after auto-retry).
Warm-start also indirectly mitigates stochastic failures by reducing
the iteration count (fewer decision points $\Rightarrow$ fewer
opportunities for the LLM to produce a non-tool-call response).
Lazy retrieval provides complementary value on simulation-error
recovery but is secondary for the reliability gain.

\section{Production Agent Quality Metrics}
\label{sec:metrics}

Beyond the controlled ablation, we analyse the system's behaviour across all
simulation runs accumulated during development, testing, and ablation
experiments.  Of these, 805 originate from an automated parameter sweep across 8 physics
studies (geometry, material, boundary conditions, initial conditions,
time-stepping, 3D geometry, Robin convection, and multi-material) run via
\texttt{scripts/sweep\_full\_study.py}; the remainder are interactive and
ablation runs accumulated through normal system use.

\begin{table}[htbp]
\centering
\caption{Agent ecosystem performance metrics from production database.
\revised{FEniCSx runs, through 2026-04-17 (excludes the controlled ablation
campaign and non-FEniCSx backends).}}
\label{tab:agent-metrics}
\begin{tabular}{lr}
\toprule
Metric & Value \\
\midrule
Total simulation runs & 1369 \\
Overall success rate & 97.8\% \\
Unique agent tasks & \revised{421} \\
First-try success rate & \revised{85.4\%} \\
\revised{Retry rate} & \revised{10.3\%} \\
Suggestion acceptance rate & 0.0\% \\
Avg.\ steps/task (analytics) & 10.2 \\
Avg.\ steps/task (database) & 6.2 \\
Avg.\ steps/task (simulation) & \revised{11.1} \\
Avg.\ simulation time & \revised{2.49s} \\
\bottomrule
\end{tabular}
\end{table}

\subsection{Success and Failure Modes}

The 97.8\% overall success rate (1{,}339/1{,}369 from PostgreSQL
\texttt{simulation\_runs}) indicates production-grade reliability
for the full agent stack.  Of the 2.2\% failures, the majority are attributable to invalid boundary condition
combinations (e.g.\ pure Neumann systems without a reference point), mesh
resolution requests exceeding GPU memory, and numerical divergence on
stiff transients.

\subsection{Iteration Efficiency}

\revised{Mean reasoning steps per completed task: Simulation Agent 11.1,
Analytics Agent 10.2, Database Agent 6.2.  The Simulation Agent's count
covers a full cycle of validating a configuration, revising it, launching
the solver and checking the result, and configurations are often revised
more than once before launch.  The Analytics Agent runs several
comparison queries before settling on a recommendation.}  The Database
Agent's lower count reflects the relative simplicity of structured
SQL-style queries over natural-language reasoning.

\subsection{First-Try Rate}

\revised{85.4\% of simulation tasks succeed on the first attempt, with no
modification-and-retry loop.  Another 10.3\% call the solver more than
once, or fall back to \texttt{debug\_simulation}, before succeeding;
these are usually boundary condition corrections or CFL stability
adjustments.  The remaining 4.3\% fail on a single attempt and never
retry.}

\revised{Attempts are counted from \texttt{run\_simulation} tool calls rather
than from distinct run identifiers.  Counting identifiers does not work
here: the agent logger back-fills the most recent \texttt{run\_id} across
every step of a task once it is known, which makes a task that retried
look identical to one that did not.  Counting tool calls recovers the
real attempt count.  Retries turn out to be less frequent than the
aggregate success rate suggests, but they are still what carries the
system from 85.4\% to 97.8\% final success.}

\subsection{KG Growth and Learning Curve}
\label{sec:kg-growth}

To isolate the KG's causal effect on performance, we run a
\emph{controlled} growth experiment.  The KG is initialised with only
static material definitions (13 materials, 45 reference entries) but
\emph{zero} Run nodes, so the agent has no prior simulation experience.
We then run 100 tasks (two shuffled passes through the 50-task
benchmark) sequentially in KG Smart mode with writes enabled; each
successful run adds one \texttt{Run} node with an HNSW-indexed
embedding and up to five \texttt{SIMILAR\_TO} edges.  By pass~2, the
KG contains 50 Run nodes from pass~1, providing warm-start context
that was absent during pass~1.

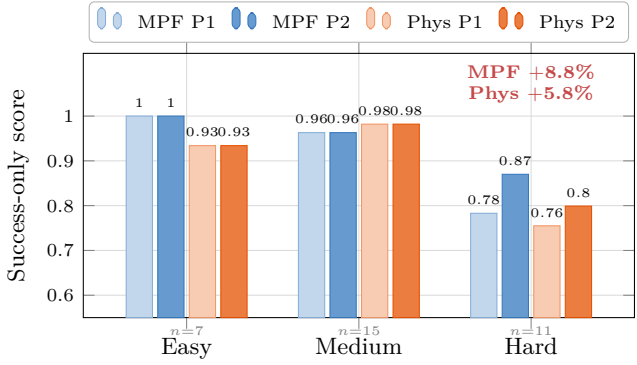
\begin{figure}[htbp]
  \centering
  \resizebox{\linewidth}{!}{
\begin{tikzpicture}
\begin{axis}[
  ybar,
  bar width=10pt,
  width=1.05\columnwidth,
  height=0.6\columnwidth,
  ylabel={Success-only score},
  ylabel style={font=\small},
  symbolic x coords={Easy, Medium, Hard},
  xtick=data,
  xticklabel style={font=\small},
  tick label style={font=\scriptsize},
  ymin=0.55, ymax=1.14,
  ytick={0.6,0.7,0.8,0.9,1.0},
  enlarge x limits=0.30,
  legend style={font=\scriptsize, at={(0.5,1.04)}, anchor=south,
                fill=white, draw=gray!60, rounded corners=2pt,
                column sep=4pt},
  legend columns=4,
  nodes near coords,
  nodes near coords align={vertical},
  nodes near coords style={font=\tiny, /pgf/number format/.cd, fixed, precision=2},
  grid=major,
  grid style={line width=0.3pt, draw=gray!30},
  clip=false,
]

\addplot[fill=nodeblue!30, draw=nodeblue!60] coordinates {
  (Easy,   1.000)
  (Medium, 0.963)
  (Hard,   0.783)
};
\addlegendentry{MPF P1}

\addplot[fill=nodeblue!75, draw=nodeblue] coordinates {
  (Easy,   1.000)
  (Medium, 0.963)
  (Hard,   0.870)
};
\addlegendentry{MPF P2}

\addplot[fill=nodeorange!30, draw=nodeorange!60] coordinates {
  (Easy,   0.934)
  (Medium, 0.982)
  (Hard,   0.755)
};
\addlegendentry{Phys P1}

\addplot[fill=nodeorange!75, draw=nodeorange] coordinates {
  (Easy,   0.934)
  (Medium, 0.982)
  (Hard,   0.799)
};
\addlegendentry{Phys P2}

\node[font=\scriptsize\bfseries, nodered!80, anchor=south, align=center]
  at (axis cs:Hard, 1.01) {MPF $+$8.8\%\\Phys $+$5.8\%};

\node[below, font=\tiny, gray] at (axis cs:Easy, 0.55) {$n{=}7$};
\node[below, font=\tiny, gray] at (axis cs:Medium, 0.55) {$n{=}15$};
\node[below, font=\tiny, gray] at (axis cs:Hard, 0.55) {$n{=}11$};

\end{axis}
\end{tikzpicture}}
  \caption{Paired KG growth comparison: success-only MPF and physics
    score for the same tasks run in pass~1 (0 prior Run nodes) vs.\
    pass~2 (50 prior Run nodes).  Only tasks succeeding in both passes
    are compared ($n{=}33$; novel tasks omitted as they rely on
    pre-seeded material definitions).  Easy and medium tasks are at
    ceiling; \textbf{hard tasks gain $+$8.8\% MPF and $+$5.8\% physics
    score}, the category where similarity-based warm-start adds most
    value.}
  \label{fig:kg-growth}
\end{figure}

To measure the effect, we perform a \emph{paired comparison}: for
each of the 50 benchmark tasks, we compare success-only MPF and
physics score between pass~1 (zero prior runs) and pass~2 (50 prior
runs), retaining only the 33 non-novel tasks that succeeded in both
passes.  Novel tasks are excluded because their fidelity depends on
pre-seeded Material nodes rather than accumulated Run history.
\cref{fig:kg-growth} shows the result stratified by difficulty.

The experiment reveals a \emph{stratified} growth effect.  Easy and
medium tasks already achieve MPF\,$\geq$\,0.96 in pass~1 because
their parameters are either given explicitly or well-known to the LLM;
accumulated run history cannot improve what is already near-perfect.
Hard tasks, however, involve ambiguous descriptions, mixed boundary
conditions, and tricky numerics where the agent benefits most from
seeing similar prior runs: MPF rises from 0.783 to 0.870 ($+$8.8\%),
and physics score from 0.755 to 0.799 ($+$5.8\%).  The overall
success rate is 88\% (88/100), confirming that KG Smart remains robust
even when starting without prior experience.

These results show that KG value is \emph{difficulty-dependent}: static
material knowledge covers simple cases, while accumulated run
experience provides measurable gains on the hardest tasks where
warm-start context disambiguates underspecified problems.

\subsection{Comparison with Prior Work}

\paragraph{FEM code generation.}
\citet{bauer2023llm} report $\approx 60\%$ one-shot success for
GPT-4-driven FEniCS script generation.  \revised{Our 85.4\% first-try rate is
higher, but the two numbers measure different tasks and are not directly
comparable: their system has to emit a complete, syntactically valid
FEniCS script, while ours fills in a parameter set for a fixed,
pre-verified solver.  What does carry across is the architecture.  Unlike}
their single-turn approach, PDE-Agents recovers from failures through
multi-turn self-correction, reaching 97.8\% final success with a fully
local, open-source model stack.
ALL-FEM \citep{deotale2026allfem} addresses the FEniCS automation problem
through fine-tuning: a corpus of 1{,}000+ verified FEniCS scripts is
used to fine-tune models from 3B to 120B parameters, and a multi-agent
framework orchestrates problem formulation, code generation, and
debugging.  Their best model (GPT~OSS 120B) achieves 71.8\% code-level
success on 39 benchmarks spanning elasticity, plasticity, fluid flow,
and multiphysics, a substantially broader PDE scope than ours.
PDE-Agents differs in three respects: (i) we use unmodified
open-source LLMs without domain-specific fine-tuning, relying instead
on tool-call orchestration and KG-augmented prompting;
(ii) our system manages the full simulation lifecycle (configuration,
execution, monitoring, debugging) rather than generating standalone
scripts; and (iii) we provide a formal V\&V study and a controlled
ablation of knowledge graph integration modes, establishing when and
why retrieval augmentation helps.

\paragraph{FEA benchmarks.}
FEABench \citep{mudur2025feabench} provides a systematic evaluation of
LLM and agent capabilities for finite element analysis using COMSOL
Multiphysics.  Their best agent strategy generates executable API calls
88\% of the time across problems in heat transfer, structural mechanics,
and electromagnetics.  Our work complements FEABench by targeting the
open-source FEniCSx ecosystem and measuring not just execution success
but \emph{output quality} (physics score, material property fidelity),
exposing cases where computationally ``successful'' simulations produce
physically incorrect results due to fabricated material properties.

\paragraph{CFD agent systems.}
The computational fluid dynamics community has been particularly active
in LLM-driven automation.  OpenFOAMGPT~2.0 \citep{chen2025openfoamgpt2}
achieves 100\% success across 450+ simulations using a four-agent
pipeline (pre-processing, prompt generation, simulation, post-processing)
with RAG-augmented retrieval.
MetaOpenFOAM \citep{chen2024metaopenfoam} decomposes CFD workflows into
subtasks using MetaGPT's \citep{hong2024metagpt} assembly-line paradigm,
reporting 85\% pass
rates at \$0.22 per case.
These systems validate the multi-agent pattern we adopt but target a
different solver ecosystem and do not evaluate knowledge graph
integration or measure output fidelity against ground-truth material
properties.

\paragraph{KG-augmented agents for materials science.}
\citet{buehler2024mechgpt} demonstrates retrieval-augmented ontological
knowledge graph strategies for materials design with a fine-tuned
MechGPT model, showing that graph-structured retrieval
outperforms flat RAG by providing mechanistic relationships between
concepts.  Our KG Smart pattern shares the same insight (that structured
knowledge improves over na\"ive retrieval) but applies it in a
\emph{live simulation loop} rather than a generative design setting,
and our ablation study provides controlled evidence for when
graph-augmented retrieval helps (novel materials) versus when it adds
overhead without benefit (standard tasks with well-known properties).

\section{System Limitations and Future Work}
\label{sec:limitations}

\paragraph{Solver scope.}
The current solver supports only scalar heat equations.  \revised{This is the main limitation of the present work.  Extending to
vector mechanics (linear elasticity, Navier--Stokes) is not simply a
matter of richer tool schemas; it requires new solver kernels, each with
its own variational form, function spaces and verification.}

\revised{The narrow scope is also what makes our central measurement
possible.  Because the agent fills in parameters for a fixed,
independently verified solver instead of emitting solver code, every run
produces a configuration whose material properties can be checked
directly against ground truth, which is what MPF (\cref{sec:novidium})
measures.  A system that generates solver code covers far broader physics
but cannot isolate property fidelity the same way, since a failure may
lie in the code, in the parameters, or in both.}

\revised{The obvious next step is to test on physics we did not write
ourselves.  ALL-FEM \citep{deotale2026allfem} has released a public
benchmark of FEM problems with ground-truth solutions covering
elasticity, plasticity, Newtonian and non-Newtonian flow, and
fluid--structure
interaction.\footnote{\url{https://github.com/fenics-llm/FEM_Dataset}}
Most of these are beyond our solver today, so evaluating on them means
implementing the corresponding kernels first.  We see that work,
combining KG-augmented memory with multi-physics breadth, as more
valuable than an incremental extension of the present study.}

\paragraph{Model scale and open-source evolution.}
The system defaults have been upgraded from the initial v1 models
(\texttt{qwen2.5-coder:32b}, \texttt{llama3.3:70b}) to
Qwen3-Coder-Next (80B total, 3B active MoE) \citep{qwen3coder2026}
and Llama~4~Scout (109B total, 17B active MoE) \citep{llama4}, with
cross-model validation confirming model-agnostic KG~Smart performance
(\cref{tab:multi-model}).  Qwen3-Coder-Next achieves higher config
quality on standard KG~Smart tasks (0.94 vs.\ 0.86) and perfect MPF on
novel tasks, identical to the v1 model.  Llama~4~Scout provides 10M-token
context and native tool calling for the orchestrator and analytics agents.
A systematic multi-model ablation across additional families (GLM-5,
DeepSeek-R2) is planned as future work.

\paragraph{KG integration.}
The KG Smart pattern (\cref{sec:ablation-disc}) achieves 100\% success
while producing the highest quality outputs, matching KG~Off's
reliability.  KG~On retains a small gap (94\%) due to budget exhaustion.
The adaptive decision framework (Algorithm~\ref{alg:kg-decision}) provides
practical deployment guidance, achieving 100\% mode-selection accuracy.
Future work will explore:
(a) reinforcement learning from simulation outcomes to tune KG retrieval
relevance, (b) dynamic confidence-based retrieval (only query KG when the
LLM's uncertainty exceeds a threshold), and (c) richer warm-start context
including failure diagnostics from similar past runs.

\revised{\paragraph{Is retrieval necessary, or would a better prompt suffice?}
A fair objection to any retrieval-augmented result is that the same
information could simply be written into the prompt.  We tested this with
a \emph{prompt-injection} control on the preliminary 7-task
novel-material benchmark: the KG was disabled and the true properties of
Novidium, Cryonite and Pyrathane were appended verbatim to each task
description.  It matched KG~Smart on fidelity (MPF~$=1.00$) and was the
fastest mode overall ($11.7$\,s mean).  That is the expected outcome,
since the answer was handed to the model.}

\revised{The control still tells us something useful about where the
difficulty lies.  Prompt injection assumes you already know which
materials a task will involve and what their properties are.  It is a
manual step, repeated per task, and it does not scale beyond a handful of
known entities.  KG~Smart reaches the same fidelity by retrieving those
properties through embedding similarity, without being told which
material the task names.  The result worth reporting is not that
retrieved context helps, which is unsurprising, but that automatic
retrieval matches hand-curated context with no human in the loop.}

\paragraph{Verification scope.}
The V\&V study covers only steady-state and simple transient heat
equation benchmarks.  A complete V\&V programme would include
higher-order elements, singularity-containing domains, and
time-dependent convergence rates.

\paragraph{Parallelism and scalability.}
The current architecture runs agents sequentially.  A task-level
parallelism layer (e.g.\ running a parametric sweep as concurrent
simulation agent calls) is under development.

\paragraph{Knowledge graph evolution.}
The 805-run parameter sweep has already created a dense graph with
4{,}000+ \texttt{SIMILAR\_TO} edges.  Future work will mine this corpus
for \texttt{IMPROVED\_OVER} relationships (did increasing mesh resolution
reduce error without proportionally increasing wall time?), enabling the
agent to autonomously suggest Pareto-optimal configurations.  A second
planned enhancement is citation traceability: agents citing specific
\texttt{ReferenceChunk} IDs from indexed literature when justifying
parameter choices, providing a fully auditable reasoning trail.

\section{Conclusion}
\label{sec:conclusion}

We have presented PDE-Agents, an open, containerised multi-agent ecosystem
that automates finite element simulations through natural-language interaction
with locally-deployed open-source LLMs.  The system achieves 97.8\% overall
success across 1{,}369 production runs and demonstrates $\mathcal{O}(h^2)$
spatial convergence validated against analytical solutions.

\paragraph{The KG Smart integration pattern.}
Our three-way ablation study across 50 benchmark tasks with a frozen KG
reveals that the \emph{integration pattern}, not the knowledge source
itself, determines KG utility.  KG~Off (no knowledge graph) achieves
100\% task completion but fabricates material properties, producing
MPF\,=\,0.34 and $\text{MPF}_w{=}0.21$ on novel materials.
KG~On (mandatory KG access) retrieves correct properties but retains
a 6\% failure rate from budget exhaustion and timeout (even after
implementing agent-level retry to eliminate stochastic failures).
KG~Smart, combining warm-start injection with lazy conditional
retrieval, achieves \textbf{100\% success with the highest output
quality} (physics 0.933, MPF 0.926), resolving the
reliability-vs-fidelity trade-off.

\paragraph{Failure analysis: why warm-start dominates.}
After implementing auto-retry to eliminate stochastic LLM failures,
KG~On's 3 remaining failures (\cref{sec:failure-analysis}) are all
systematic: 2 budget-exhaustion cases and 1 timeout, all caused by
the mandatory pre-simulation KG query loop.  Warm-start injection
eliminates this bottleneck entirely by providing material properties
and reference configurations before the agent loop begins,
accounting for KG~Smart's 6-pp success rate advantage.  Lazy retrieval
provides complementary value on failure recovery but is secondary
for the reliability gain.

\paragraph{KG value is difficulty- and domain-dependent.}
A controlled 100-task KG growth experiment shows that the KG's benefit
is concentrated where it matters most.  Easy and medium tasks are already
at quality ceiling ($\text{MPF}{\geq}0.96$) from the first pass; hard
tasks, involving ambiguous descriptions and mixed boundary conditions, gain
$+$8.8\% MPF and $+$5.8\% physics score from accumulated run history.
For novel materials, the KG is indispensable: fabricated properties cause
output errors up to 949\,K, while KG~Smart eliminates these errors entirely.

\paragraph{Lessons learned.}
Three design principles emerged from the ablation and failure analysis:
\begin{enumerate}
  \item \textbf{Never make KG access mandatory in the agent's critical path.}
        Mandatory pre-simulation retrieval creates fragile coupling between
        knowledge access and task completion.  Front-loading context via
        embedding similarity (warm-start) decouples them.
  \item \textbf{Front-load context, don't force the agent to query for it.}
        The agent's iteration budget is its scarcest resource.  Injecting
        relevant context into the system prompt before the agent loop is
        strictly more efficient than requiring the agent to spend iterations
        formulating and executing KG queries.
  \item \textbf{Let the agent decide when it needs more knowledge.}
        Lazy conditional retrieval respects the agent's autonomy: it queries
        the KG only after a failure or when material properties are genuinely
        unknown, avoiding unnecessary round-trips on tasks where the LLM's
        parametric knowledge suffices.
\end{enumerate}

\paragraph{Broader implications.}
The warm-start + lazy-retrieval pattern is not specific to heat transfer
simulation.  Any tool-using agent domain with a growing knowledge base
(materials science, drug discovery, circuit design, structural
engineering) could benefit from the same architecture: embed the task,
retrieve the most similar prior successes, inject them as few-shot
context, and defer explicit knowledge queries to the failure-recovery
path.  As knowledge graphs grow and LLMs improve, the quality gap
between KG-enabled and KG-free agents will only widen for novel or
proprietary domains where the LLM's training data provides no coverage.

The production metrics confirm that multi-turn self-correction is
essential: the \revised{85.4\%} first-try rate improves to 97.8\% through
iterative debugging, underscoring the value of agent-level orchestration
over single-turn code generation.  PDE-Agents establishes both a working
platform and a rigorous empirical baseline for the emerging field of
LLM-driven autonomous simulation, showing that structured knowledge
integration turns an agent from an ``expensive random number
generator'' into a reliable engineering tool.

\section*{Data Availability}
All code, evaluation scripts, and result JSON files are available at
\url{https://github.com/MatPro-IFE/pde-agents}.  The Neo4j knowledge
graph seed data is included in the repository at \texttt{knowledge\_graph/seeder.py}.

\section*{Author Contributions}
\textbf{Sayan Adhikari}: Conceptualization, Methodology, Investigation, Software, Data Curation, Formal Analysis, Visualization, Writing -- Original Draft, Writing -- Review \& Editing.
\textbf{Gulshan Noorsumar}: Conceptualization, Methodology, Investigation, Validation, Writing -- Review \& Editing.
\textbf{\O yvind Jensen}: Supervision, Writing -- Review \& Editing.

\section*{Declaration of Generative AI in the Writing Process}
During the preparation of this manuscript, Anthropic's Claude Opus~4.6 was used to assist with language editing, grammar refinement, and the programmatic generation of TikZ figures (Figures~1 and~2).
All scientific content, experimental design, analysis, and interpretation were performed by the human authors, who reviewed and take full responsibility for the final manuscript.

\section*{Competing Interests}
The authors declare no competing interests.

\section*{Acknowledgements}
The authors thank the anonymous reviewers for their constructive feedback.
Computational resources were provided by the Institute for Energy Technology (IFE).

\bibliographystyle{unsrtnat}
\bibliography{references}

\appendix

\section{\texorpdfstring{\revised{Full h-Refinement Data}}{Full h-Refinement Data}}
\label{app:vv-detail}

\revised{\Cref{tab:vv-detail} reports the full mesh-refinement study behind the
convergence rates summarised in \cref{tab:vv-convergence}: $L^2$ and
$L^\infty$ error norms at every resolution $N \in \{8, 16, 32, 64, 128\}$
for all three benchmark cases.}

\revised{The two norms are computed differently, and for one case they disagree
by several orders of magnitude.  $\|e\|_{L^2}$ is assembled by quadrature
over each element, so it measures error throughout the domain, including
between nodes.  $\|e\|_{L^\infty}$ is evaluated only at the degrees of
freedom.  For the constant-source (1D-like Poisson) case the P1 solution
is \emph{nodally exact}, which is a standard superconvergence result, so
the nodal norm sits at machine precision ($\approx 10^{-10}$) at every
resolution while the integrated norm decays at $\mathcal{O}(h^2)$ as
expected.  Convergence rates are therefore fitted from $\|e\|_{L^2}$
alone; a rate fitted from nodal values would say nothing useful about
this case.}

\begin{table*}[htbp]
\centering\footnotesize
\caption{Detailed convergence data: error norms at each mesh resolution.  $\|e\|_{L^2}$ is integrated over each element with quadrature of degree 8, so it captures the error \emph{between} nodes; $\|e\|_{L^\infty}$ is evaluated at the degrees of freedom only.  For the constant-source case the P1 solution is nodally exact, so $\|e\|_{L^\infty}$ sits at machine precision while $\|e\|_{L^2}$ shows the expected $\mathcal{O}(h^2)$ interpolation error.  The two norms measure different things, and only the integrated norm is used to fit convergence rates.}
\label{tab:vv-detail}
\begin{tabular}{lrrrr}
\toprule
Case & $N$ & DOFs & $\|e\|_{L^2}$ & $\|e\|_{L^\infty}$ \\
\midrule
  2D Steady-State Linear Profile & 8 & 81 & 4.22e-11 & 1.12e-10 \\
   & 16 & 289 & 3.65e-09 & 1.44e-08 \\
   & 32 & 1,089 & 3.89e-10 & 1.11e-09 \\
   & 64 & 4,225 & 1.10e-09 & 3.00e-09 \\
   & 128 & 16,641 & 2.47e-09 & 6.40e-09 \\
\addlinespace
  2D Transient Fourier Mode Decay & 8 & 81 & 4.38e-03 & 4.88e-03 \\
   & 16 & 289 & 9.48e-04 & 2.64e-03 \\
   & 32 & 1,089 & 2.39e-04 & 6.62e-04 \\
   & 64 & 4,225 & 5.98e-05 & 1.65e-04 \\
   & 128 & 16,641 & 1.50e-05 & 4.13e-05 \\
\addlinespace
  2D Steady-State with Constant Source (1D-like) & 8 & 81 & 1.43e-03 & 1.12e-10 \\
   & 16 & 289 & 3.57e-04 & 2.28e-11 \\
   & 32 & 1,089 & 8.91e-05 & 1.22e-10 \\
   & 64 & 4,225 & 2.23e-05 & 1.83e-10 \\
   & 128 & 16,641 & 5.57e-06 & 2.68e-10 \\
\addlinespace
\bottomrule
\end{tabular}
\end{table*}

\end{document}